\documentclass[pra,twocolumn,showpacs,preprintnumbers,superscriptaddress,
amsmath,amssymb,tightenlines,epsfig,floatfix]{revtex4}

\usepackage{bm}
\usepackage{graphicx}
\usepackage{color}

\newcommand{\beq}{\begin{equation}}
\newcommand{\eeq}{\end{equation}}

\newcommand{\h}{\hat}

\newcommand{\be}{\begin{equation}}
\newcommand{\ee}{\end{equation}}
\newcommand{\bi}{\begin{itemize}}
\newcommand{\ei}{\end{itemize}}
\newcommand{\ben}{\begin{enumerate}}
\newcommand{\een}{\end{enumerate}}

\newcommand{\rv}{{\bf r}}

\newcommand{\bea}{\begin{eqnarray}}
\newcommand{\eea}{\end{eqnarray}}
\newcommand{\up}{\uparrow}
\newcommand{\down}{\downarrow}
\newcommand{\<}{\langle}
\renewcommand{\>}{\rangle}
\renewcommand{\(}{\left(}
\renewcommand{\)}{\right)}
\renewcommand{\[}{\left[}
\renewcommand{\]}{\right]}
\newcommand{\commentout}[1]{{}}
\newcommand{\half}{\hbox{$1\over2$}}
\newcommand{\eq}[1]{Eq.~\eqref{#1}}

\definecolor{black}{rgb}{0,0,0}

\definecolor{blue}{rgb}{0,0,1}

\definecolor{green}{rgb}{0,1,0}

\definecolor{red}{rgb}{1,0,0}

\begin{document}
\title{Manipulating atoms in an optical lattice:
Fractional fermion number\\ and its optical quantum measurement}
\author{J. Ruostekoski}\email{janne@soton.ac.uk}
\affiliation{School of Mathematics, University of Southampton,
Southampton, SO17 1BJ, UK}
\author{J. Javanainen}\email{jj@phys.uconn.edu}
\affiliation{Department of Physics, University of Connecticut,
Storrs, CT 06269, USA}
\author{G. V. Dunne}\email{dunne@phys.uconn.edu}
\affiliation{Department of Physics, University of Connecticut,
Storrs, CT 06269, USA} \affiliation{CSSM, Department of Physics,
University of Adelaide, SA 5005, Australia} \affiliation{Institut
f\"ur Theoretische Physik, Universit\"at Heidelberg, Philosophenweg
16, 69120 Heidelberg, Germany}

\begin{abstract}
We provide a detailed analysis of our previously proposed scheme
[Phys.\ Rev.\ Lett.\ {\bf 88}, 180401, (2002)] to engineer the
profile of the hopping amplitudes for atomic gases in a 1D optical
lattice so that the particle number becomes fractional. We consider
a constructed system of a dilute two-species gas of fermionic atoms
where the two components are coupled via a coherent electromagnetic
field with a topologically nontrivial phase profile. We show both
analytically and numerically how the resulting atomic Hamiltonian in
a prepared dimerized optical lattice with a defect in the pattern of
alternating hopping amplitudes exhibits a fractional fermion number.
In particular, in the low-energy limit we demonstrate the
equivalence of the atomic Hamiltonian to a relativistic Dirac
Hamiltonian describing fractionalization in quantum field theory.
Expanding on our earlier argument [Phys.\ Rev.\ Lett.\ {\bf 91},
150404 (2003)] we show how the fractional eigenvalues of the
particle number operator can be detected via light scattering. In
particular, we show how scattering of far-off resonant light can
convey information about the counting statistics of the atoms in an
optical lattice, including state-selective atom density profiles and
atom number fluctuations. Optical detection could provide a truly
quantum mechanical measurement of the particle number
fractionalization in a dilute atomic gas.
\end{abstract}

\date{\today}
\maketitle

\section{Introduction}

Loading alkali-metal atoms into optical lattices has opened up
fascinating possibilities to study many-particle systems with a wide
range of technology to manipulate the parameters of the system
Hamiltonian as well as the quantum states of the atoms. In
particular, the flexibility of the experimental preparation of
ultra-cold atoms could be an advantage in realizing atomic systems
with close analogies to the physical systems, known in other
disciplines of modern physics, in which the experimental evidence or
theoretical understanding is limited.

We previously proposed a scheme to engineer the spatial profiles of
the hopping amplitudes for atoms in an optical lattice \cite{RUO02}.
This is achieved by superimposing coherent electromagnetic
(em) field amplitudes that induce a coupling between different spin
states of the atom and drive transitions between neighboring lattice
sites. We considered a particular constructed system of a
two-species gas of fermionic atoms in a 1D lattice, where the two
components are coupled via an em field with a topologically
nontrivial phase profile (with topological properties similar to a
soliton, or a phase kink), so that the particle number becomes
fractional. In particular, we showed that, in the low energy limit,
the resulting Hamiltonian of the atomic system has a one-to-one
correspondence to a relativistic Dirac Hamiltonian describing
fractional fermions in quantum field theory \cite{JAC76,NIE86,jackiw}. This
is related to fractionalization in the polyacetylene polymer systems,
where the linearized lattice vibrations are coupled to the electrons
so that the dynamics of the electrons is described by a relativistic
Dirac equation. We also showed that the fractional part of the
particle number in the atomic system can be accurately controlled by
modifying the effective detuning of the em field.

Moreover, we proposed an optical measurement scheme for fractional
particle number in an atomic gas that could detect directly the
fractional eigenvalue and determine atom number fluctuations \cite{JAV03}.
   In particular, the fractional fermion could be imaged in
a phase-contrast set-up, while the atom number fluctuations are
reflected in the fluctuations of the number of photons in the
light scattered off the lattice.

In this paper we present a detailed description of how
fractionalization manifests itself and how it could be optically
detected in the atomic regime, using an optically trapped
Fermi-Dirac (FD) atomic gas. We show how to engineer spatially
inhomogeneous hopping amplitude profile for the atoms in an optical
lattice. We extend our analysis in Ref.~\cite{RUO02} of different
optical and rf/microwave fields to prepare first a dimerized lattice
and later to include a discontinuity in the hopping amplitude
pattern. We demonstrate both analytically and numerically how the
resulting atomic Hamiltonian exhibits a fractional fermion number.
We consider different observables, accessible via light scattering
experiments, that could directly probe the fractional fermion number
and its fluctuations.

Adjusting the hopping amplitudes in our scheme in Ref.~\cite{RUO02}
provided the means of preparing a topologically non-trivial atomic
ground state. More recently, similar techniques to engineer the
spatial profile of the hopping amplitude between the atoms in
adjacent optical lattice sites using em transitions were proposed to
create effective magnetic fields in a 2D lattice \cite{JAK03,MUE04},
as well as non-Abelian gauge fields for the atoms in an optical
lattice \cite{OST05}. In Ref.~\cite{JAK03} a spatially alternating
pattern of hopping amplitudes along one spatial direction was used
to induce a non-vanishing phase of particles moving along a closed
path on the lattice. Since this phase is proportional to the
enclosed area, the hopping amplitude pattern simulates the effects of
a magnetic field on charged particles. By
introducing a level degeneracy in such a set-up allowed to consider
non-Abelian phases for the atoms moving along closed paths
\cite{OST05}.

An interesting property of our constructed optical lattice
Hamiltonian is that the ultra-cold fermionic atoms exhibit
long-wavelength dynamics that is equivalent to relativistic Dirac
fermions with a controllable mass and with coupling to a bosonic
field. Here the bosonic field is represented by the external em
field amplitude inducing the hopping of the atoms between adjacent
lattice sites and the mass of the Dirac fermions is proportional to
the effective detuning of the coupling field from the resonance. The
realizations of emerging relativistic Dirac quasiparticles in
condensed matter systems has recently attracted considerable
interest, e.g., in graphene systems where the 2D hexagonal lattice
structure has low-energy excitations analogous to those of Dirac
fermions \cite{NOV05,ZHA05,KAT06,ZHU07}. The atomic optical lattice
systems can be engineered and controlled to a much higher degree
than graphene or polymer systems. For instance, the atomic gas could
be used to demonstrate the effect of the Dirac fermion mass on the
fractional part of the eigenvalue \cite{GW} and the
finite-temperature contributions to the fractional fermion number in
relativistic field theory \cite{AD,DR}. The optical lattice could
provide a laboratory to simulate the properties of relativistic
Dirac fermions also more generally.

Particle number fractionalization is a remarkable phenomenon in both
relativistic quantum field theory and condensed matter systems
\cite{jackiw,anderson}. Jackiw and Rebbi \cite{JAC76,NIE86} showed
that for a fermionic field coupled to a bosonic field with a
topologically nontrivial soliton profile, the fermion number can be
fractional. The noninteger particle number eigenvalues may be
understood in terms of the deformations of the Dirac sea (or the
hole sea) due to its interaction with the topologically nontrivial
environment. Fractional fermion number has been discussed
previously in the condensed matter regime in 1D conjugated polymers
(the SSH model) \cite{SU79,HEE88,NIE86}. The existence of
fractionally charged excitations in the polymers is typically
demonstrated indirectly by detecting the reversed spin-charge
relation~\cite{HEE88}. The fractional quantum Hall effect (FQHE) can
also be explained by invoking quasiparticles, each with a fraction
of an electron's charge~\cite{laughlin}, but the fractionalization
mechanism is very different from that in the polymers and in the
atomic gas in this paper. The fluctuations of the tunneling current
in low-temperature FQHE regime have been
measured~\cite{DEP97,SAM97}. Interpreting the current shot noise
according to the Johnson-Nyquist formula duly suggests that the
current is carried by the fractional Laughlin quasiparticles.
Analogous experiments have determined the fractional expectation
value of the charge in FQHE in the Coulomb blockade
regime~\cite{GOL95}. Recently fractional statistics was observed by
realizing a quasiparticle interferometer in the FQHE regime
\cite{CAM05}.

Light scattering can be an efficient probe of quantum statistical
properties of ultra-cold atoms. It can provide information about
spatial correlation functions between the atoms \cite{MOR95} and
spectral properties about excitations \cite{JAV95}. The scattering
process could also generate an entanglement between the photons and
the many-particle atomic state \cite{RUO97d,RUO98a}. In
Ref.~\cite{JAV03} we adapted the techniques of far off-resonant
light scattering in optically thin samples \cite{JAV95} to the
atomic gases trapped in optical lattices. In the measurements of the
expectation value of the fractional particle number, we make use of
the fact that the adjacent lattice sites are occupied by different
fermionic species. In a phase-contrast imaging, we may adjust the
light scattered from one species to add to the incident beam and the
light scattered from the other species to subtract. By measuring the
intensity of the scattered light, one may detect the atom number
fluctuations in order to demonstrate that not only the expectation
value is fractional but also that the fluctuations become
vanishingly small. This is required to show that the fractional
fermion number represents an eigenstate of a particle number
operator. The detection of both the expectation value and the
fluctuations would represent a truly quantum mechanical measurement
of fractionalization that has been a long-standing challenge in
physics. Our proposed measurement techniques could also be useful in
detecting fractional Laughlin quasi-particles in rapidly rotating
atomic gases \cite{COO01}. Although the emphasis of our optical
detection analysis was on fractionalization, the basic formalism
could also be used as a more general optical probing technique of
atom number fluctuations in optical lattices. The optical detection
of atom number fluctuations in an optical lattice using an analogous
approach and the enhancement of the signal by an optical cavity was
more recently addressed in Ref.~\cite{MEK07,MEK07b}.

In addition to providing atomic physics technology for a potentially
direct detection of the fractional fermion number and its
fluctuations that have so far escaped experimental observation, our
dilute atomic gas  has a possible advantage, compared to
condensed matter systems, in the sense that the atoms form a very
clean system: The interatomic interactions are weak and
well-understood, the lattice may be prepared without imperfections,
and different parameters of the system Hamiltonian can be
controlled.

In Sec.~\ref{optlatticesys} we introduce the basic optical lattice
system to produce a fractional fermion number. In
Sec.~\ref{basicsys} we first explain the lattice Hamiltonian and
em-induced atom hopping from site to site, then discuss the
nonuniform hopping amplitudes between adjacent lattice sites in a
dimerized lattice. Introducing a defect in the regular pattern of
hopping amplitudes results in a bound state and a fractional fermion
number, localized around the defect; methods to prepare such a
defect are also discussed. Pertinent mathematical properties of the
fractional fermion number and its fluctuations in a finite lattice
are derived in Sec.~\ref{finitesec}. An example hopping amplitude
profile that generates a fractional number 1/3 is introduced in
Sec.~\ref{onethreesec}. We show that the long-wavelength limit of
the atomic lattice Hamiltonian is equivalent to the relativistic
Dirac Hamiltonian in Sec.~\ref{relsec}. We review some basic
properties of the corresponding Jackiw-Rebbi relativistic model that
provides a simple description of the fractionalization process. The
optical detection method is explained in Sec.~\ref{optsec}. We first
show how the fractional fermion number could be imaged in a
phase-contrast set-up and then how the fluctuations of the
fractional observable are reflected in the fluctuations of
incoherently scattered light. Finally, a few concluding remarks are
made in Sec.~\ref{concsec}, where we also address the effects of the
harmonic trapping potential.

\section{Optical lattice system}\label{optlatticesys}

\subsection{Basic physical system}\label{basicsys}

In our scheme to realize particle number fractionalization in atomic
gases we consider neutral FD atoms loaded in a periodic 1D optical
lattice. The optical potential is induced by means of the
ac Stark effect of off-resonant laser beams \cite{RAI97}.

Atomic systems in 1D may be created by confining the atoms tightly
along the transverse directions. This is typically obtained by means
of optical lattices or atom chips. For instance, in recent
experiments 1D optical lattices were created from 3D optical
lattices by increasing the optical lattice barrier height along two
orthogonal directions \cite{tonks,FER05}. This resulted in a 2D
array of decoupled 1D tubes of atoms where the atoms in each tube
experienced a periodic optical standing wave potential along the
axial direction of the tube. If bosonic atoms are confined
sufficiently tightly along the two transverse directions, the system
approaches the Tonks-Girardeau regime where the impenetrable bosons
obey the FD statistics \cite{tonks}.
\begin{figure}
\includegraphics[width=0.49\columnwidth]{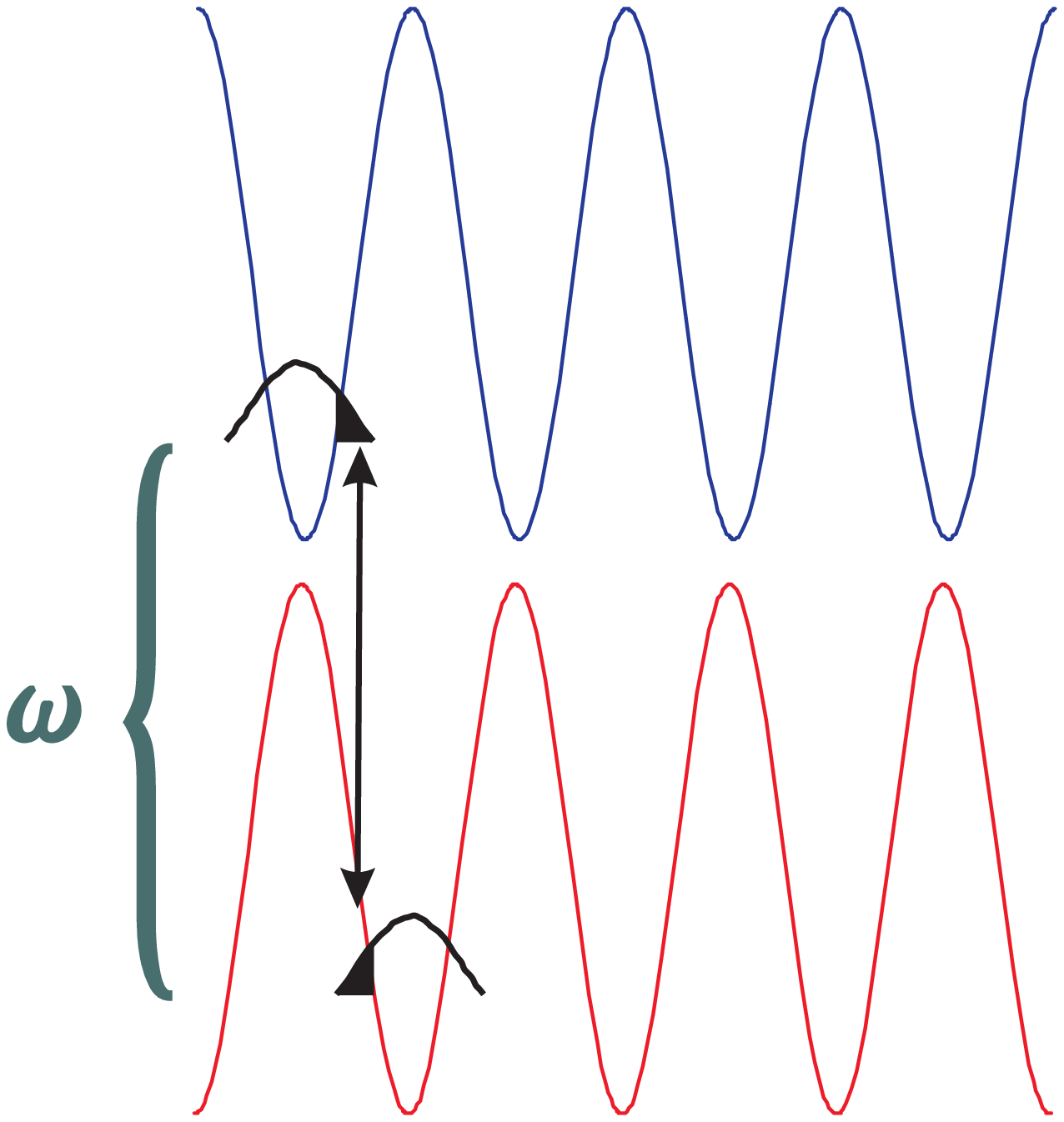}
\includegraphics[width=0.49\columnwidth]{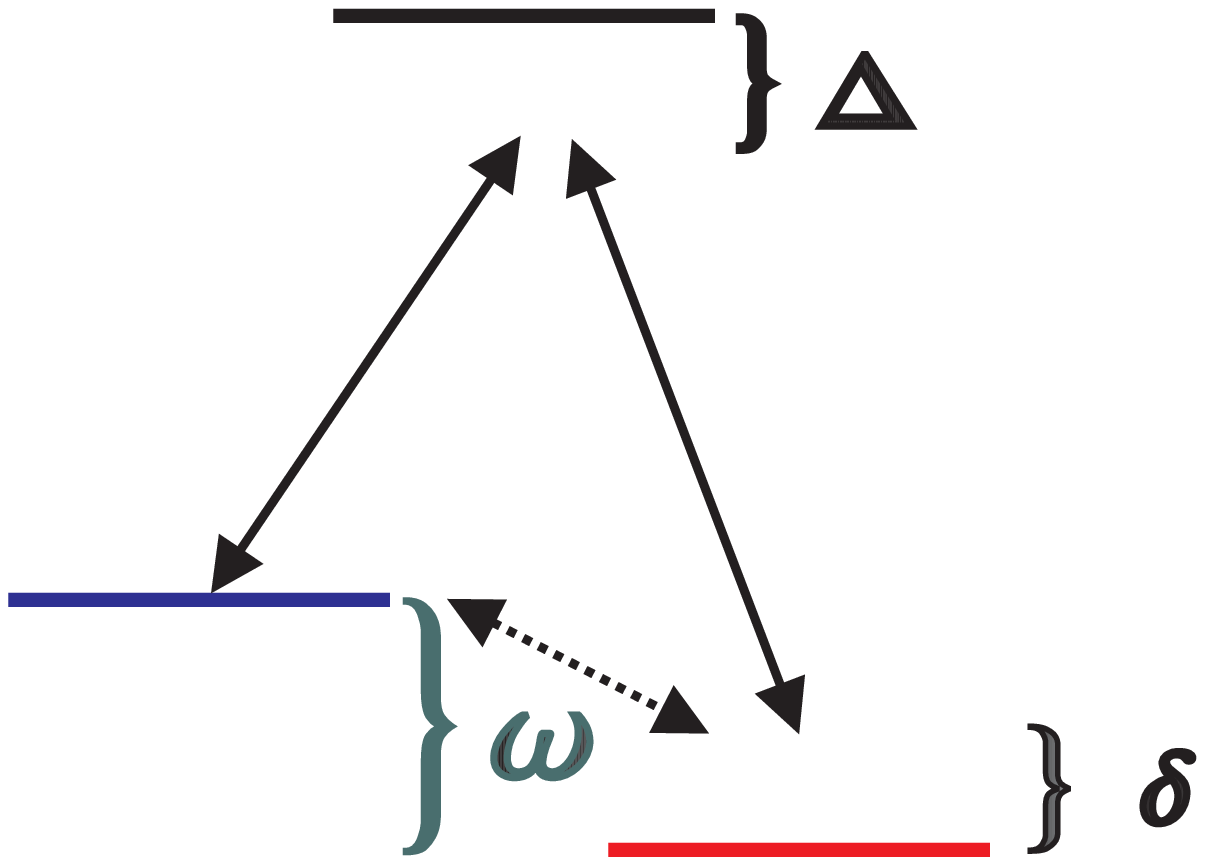}
\vspace{-3mm} \caption{The energy diagram of the two-species
Fermi-Dirac gas in an optical lattice. The atoms occupy two
different internal levels and experience different periodic optical
potentials shifted by $d$ (on the left). The coupling between the
neighboring lattice sites, representing different spin components,
is induced by two-photon Raman transitions, or by a superposition of
the Raman transition and a one-photon microwave or rf transition
(dotted line). The transition region is denoted by the dark shaded
area representing the overlap region of the atomic wave functions.
The energy difference between the two components in the overlap
region is denoted by $\omega$. The two-photon transition is
far-detuned by $\Delta$ from an intermediate atomic level (on the
right), and in this illustration $\delta$ stands for the two-photon
detuning.} \label{f1}
\end{figure}

We analyze an optical lattice system confining two fermionic species
using an atomic Hubbard Hamiltonian \cite{JAK98}. In a sufficiently
deep lattice each site is assumed to support one mode function (the
Wannier function) that is weakly coupled to two nearest neighbor
sites. We assume that the hopping of the atoms between adjacent
lattice sites only occurs as a result of driving by coherent em
fields. This could be realized, e.g., with a FD gas occupying two
internal levels $|\up\>$ and $|\down\>$ that are coherently coupled
via an em-induced transition, if the two species experience optical
potentials which are shifted relative to each other by $d$, where
$2d$ denotes the lattice period for each state $|\up\>$ and
$|\down\>$; see Fig.~\ref{f1}. The coupling could be a far
off-resonant optical Raman transition via an intermediate atomic
level, a microwave, or a rf transition. The optical lattice
potential with alternating spin states in alternating sites may be
realized, e.g., by combining $\sigma^+$ and $\sigma^-$ polarized
lasers \cite{MAN03}, or when the light beam is blue-detuned from an
internal transition of the atoms in level $|\up\>$ and red-detuned
from a transition in level $|\down\>$. A simple example 1D lattice
potential in that case is $V_\up (x) = V_0\sin^2(\pi x/2d)$, and
$V_\down (x) = -V_0\sin^2(\pi x/2d)$.

In our example set-up of Fig.~\ref{f1}, the neighboring lattice
sites represent atoms in different internal levels and are separated
by a distance $d$. Here $d=\lambda/[4\sin(\theta/2)]$ is obtained by
two laser beams, with the wavelength $\lambda$, intersecting at an
angle $\theta$. The lattice spacing can be easily modified by
changing the angle between the lasers. We may write the annihilation
operators for fermionic atoms as $c_{2k+1,\up}$ and $c_{2k,\down}$
at odd and even numbered lattice sites, occupying the internal
levels $\up$ and $\down$, respectively. In the following we suppress
the explicit reference to the different internal levels and write
the annihilation operator for the atoms in the $k$th site simply as
$c_k$. The Hamiltonian for the atomic system then reads
\begin{equation}
      \frac{H}{\hbar} = \sum_k \left[\epsilon_k c^\dagger_kc_k
-(\kappa_k c^\dagger_{k+1}c_k + \kappa_k^* c^\dagger_k
c_{k+1})\right]\,. \label{ham}
\end{equation}
The em-induced coupling between the two internal states between the
$k$th and the $(k+1)$th site is described by
\begin{equation}
\kappa_k=\int d^3 r\, \psi^*_\sigma(\rv-\rv_k)
\Omega(\rv)\psi_{-\sigma}(\rv-\rv_{k+ 1})\,, \label{kappa}
\end{equation}
and $\epsilon_k$ stands for the effective detuning and/or external
potential variation between different sites. The em-coupling terms
are the analogues of the hopping terms in the corresponding polymer
Hamiltonian \cite{NIE86}. The mode functions of the individual
lattice sites (Wannier functions) are denoted by $\psi_\sigma
(\rv-\rv_i)$, with $\sigma=\up,\down$ and $-\sigma=\down,\up$,
respectively. We assume that the em coupling between the internal
levels $\Omega(\rv)$ is the only transition mechanism
for the atoms between neighboring lattice sites, and therefore
ignore the direct tunneling. $\Omega({\bf r})$ could,
for instance, be an effective two-photon Rabi frequency at the position
${\bf r}$. In this respect our optical-lattice scheme is closely related
to the microtrap scheme discussed recently as a method of transporting
atoms for the purposes of quantum information processing~\cite{DEB07}.  We
neglect the $s$-wave scattering between the two FD species. This process
can be made weak by having a small spatial overlap between the mode
functions of the adjacent lattice sites.

Even though the optical lattice may be part of a larger trap, which
could generate interesting physics in its own right, we simplify by
setting $\epsilon_k=(-1)^k \delta/2$ to be constant along the
lattice separately for the even ($\down$) and for the odd ($\up$)
sites that are split by the energy difference $\delta$. We mostly
concentrate on the simplest case $\delta=0$. The fractional fermion
number arises from certain types of defects in the couplings
$\kappa_k$.  In this paper we illustrate by concentrating on one
such set-up; we consider a {\it dimerized} lattice generated by the
coupling matrix element that alternates from site to site between
two values $a+\mu$ and $a-\mu$, except that at the center of the
lattice there is a defect such that the same coupling matrix element
appears twice. In the following, we use the notation introduced in
Ref.~\cite{JAV03,signcomment}.

\subsubsection{Dimerized optical lattice}

In this section we introduce an alternating pattern in the atomic
hopping amplitudes. This establishes a dimerized optical lattice. If
we also add a defect in the pattern of alternating coupling matrix
elements, this results in a fractional fermion number, localized
around the defect. This type of schemes are the subject of the
following two sections.

We consider the coupling frequency $\Omega(\rv)$ to be a
phase-coherent superposition of two field amplitudes:
\begin{equation}
\Omega(\rv)= {\cal V}(\rv)+ {\cal U}(\rv)\sin\( \frac{\pi x}{d}\)
\,. \label{dime}
\end{equation}
We can assume that the matrix elements
\begin{align}
a_k &\equiv \int d^3 r\,\psi^*_\sigma (\rv-\rv_k){\cal V}(\rv)
\psi_{-\sigma} (\rv-\rv_{k + 1}),\\
\mu_k &\equiv \int d^3 r\,\psi^*_\sigma(\rv-\rv_k){\cal U}(\rv)
\psi_{-\sigma} (\rv-\rv_{k + 1})\,,
\end{align}
do not change sign over the length of the lattice.  The
purpose of the standing wave in Eq.~{(\ref{dime})} is to introduce a
spatially alternating sign for the hopping amplitude along the
lattice.

\begin{figure}
\includegraphics[width=0.95\columnwidth]{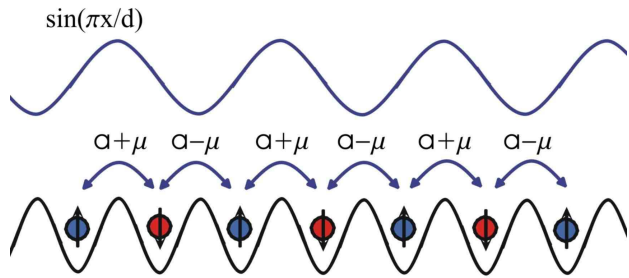}
\vspace{-3mm} \caption{Dimerized optical lattice with spatially
alternating hopping amplitude for atoms between adjacent lattice
sites. The hopping is induced by electromagnetic field (as in
Fig.~\ref{f1}) whose amplitude varies according to Eqs.~\eqref{dime}
and~\eqref{ome2}. Here $a_k$ and $\mu_k$ are assumed to be constant
along the lattice. The coefficient in the front of $\mu$ is
determined by the value of $\sin(\pi x/d)$ at the middle between the
lattice sites where the Wannier functions of the two lattice sites
overlap.
     } \label{f2}
\end{figure}

We assume that $\sin(\pi x/d)$ in Eq.~{(\ref{dime})} is
approximately constant over the small spatial overlap area of two
neighboring lattice site atom wavefunctions. One of the lattice
sites has the potential minimum at $x=0$ and its two nearest
neighbor sites minima at $x=\pm d$. The spatial overlap of the
atomic wavefunctions centered at the sites $x=0$ and $x=\pm d$ has
the maximum at $x=\pm d/2$. The $\sin(\pi x/d)$ term in the hopping
amplitude in the overlap region at $x=d/2$ is equal to one and at
$x=-d/2$ it is equal to $-1$. With similar arguments it is easy to
see that the hopping amplitude alternates between the values $\pm 1$
at each overlap area and we obtain (see Fig.~\ref{f2})
\begin{equation}
\kappa_k \simeq a_k +(-1)^k\mu_k \,. \label{ome2}
\end{equation}
In this approximation, we take $a_k$ and $\mu_k$ real and in the actual
calculations assume them to be constant along the lattice, $a_k=a$
and $\mu_k=\mu$.

The field amplitude, \eq{dime}, required to generate the hopping
amplitude, \eq{ome2}, represents a phase-coherent superposition of a
standing wave and a field amplitude with a uniform phase profile.
The field superposition could be produced, e.g., by using optical
holograms. We will address such techniques in more detail in the next
section, when we consider a more complicated field profile needed to
prepare fractional fermions. Due to the relatively simple form of
\eq{dime}, the field configuration could be generated by superposing
an optical standing wave and a rf, microwave, or an optical field
${\cal V}$. The two fields both need to couple the states $|\up\>$
and $|\down\>$, or to form part of a multi-level, multi-photon
transition that couples the two states together.

The standing wave in the hopping amplitude has the period $2d$ when
the period of the amplitude of the lattice lasers forming the
optical potential is equal to $4d$. The different periodicity may be
obtained by using a different angle $\theta'$ between two
counter-propagating lasers, so that $d=\lambda'/[2\sin(\theta'/2)]$,
where $\lambda'$ is the wavelength of the $\sin(\pi x/d)$ hopping
field.

Alternatively, the $\sin(\pi x/d)$ hopping term may be prepared by
using the same wavelength $\lambda'=\lambda$ and intersection angle
$\theta'=\theta$ as in the optical lattice potential with a
two-photon optical Raman transition (Fig.~\ref{f1}). The strength of
an off-resonant two-photon Rabi frequency in the limit of large
detuning, $\Delta$, from the intermediate state is
$\Omega\propto{\cal R}_1 {\cal R}_2/\Delta$, where ${\cal R}_i$
denote the Rabi frequencies in the individual transitions
\cite{JAV95}. For two standing-wave one-photon couplings displaced
from one another by $d$, the two-photon Rabi frequency is
$\Omega\propto\sin(\pi x/2d)\cos(\pi x/2d)\propto\sin(\pi x/d)$, so,
even though the period of the lattice laser amplitude is $4d$, the
effective field amplitude of the two-photon transition has the
desired period $2d$.

In order to understand the effect of dimerization, we follow \cite{SU79} and
consider the Hamiltonian \eqref{ham} for the uniform lattice
$\epsilon_k=0$ with the coupling \eqref{ome2} in the case of
periodic boundary conditions.
The adjacent lattice sites are separated by $d$
and we can obtain the reduced zone representation for $-\pi/2d<p\leq
\pi/2d$ by transforming the lattice operators according to
\begin{align}
c_{p-} &= \frac{1}{\sqrt{N}} \sum_k e^{ipkd} c_k,\label{va}\\
c_{p+}&= \frac{1}{ \sqrt{N}} \sum_k e^{ipkd}(-1)^k c_k\,.\label{co}
\end{align}
In terms of the valence and the conduction band operators of the
undimerized system, $c_{p-}$ and $c_{p+}$, the Hamiltonian
\eqref{ham} reads:
\begin{align}
\frac{H}{\hbar}= \sum_p& \big[ e_p(c_{p+}^\dagger c_{p+}-
c_{p-}^\dagger c_{p-})\nonumber\\ &+\Delta_p (c_{p+}^\dagger
c_{p-}+c_{p-}^\dagger c_{p+})\big] \,,
\end{align}
with
\begin{equation}
\Delta_p \equiv 2\mu \sin\({pd}\),\quad e_p \equiv 2 a
\cos\({pd}\)\,.
\end{equation}
This can be diagonalized by the Bogoliubov transformation
\begin{align}
b_{p-} &= \alpha_p c_{p-}- \beta_p c_{p+},\\
b_{p+} &= \alpha_p c_{p+}+ \beta_p c_{p-}\,,
\end{align}
with the appropriate choice of the coefficients $\alpha_p$ and
$\beta_p$, so that
\begin{equation}
\frac{H}{\hbar}= \sum_p E_p (b_{p+}^\dagger b_{p+}- b_{p-}^\dagger
b_{p-})\,.
\end{equation}
The quasiparticle energies are given by
\begin{equation}
E_p\equiv \sqrt{ e_p^2 +\Delta_p^2}\,,\label{quasip}
\end{equation}
and the spectrum exhibits an energy gap equal to $4\hbar |\mu|$ at
the reduced zone boundary $p=\pm\pi/2d$, due to the dimerization
field. We may define the atomic correlation length according to
\eq{quasip}, analogously to the BCS theory:
\begin{equation}
\xi \simeq \left| \frac{e'(p_F)}{\Delta(p_F)}\right|= \left|
\frac{ad}{\mu}\right| \,,\label{correlation}
\end{equation}
where $p_F=\pi/2d$ is the Fermi momentum.

\subsubsection{Defect in the pattern of alternating hopping
amplitudes}

In this section we construct a field amplitude that breaks the
symmetry of the dimerized lattice formed by the alternating pattern
of the hopping amplitudes. This is obtained by generating a defect
at the center of the lattice, so that the same coupling term between
the lattice sites appears twice in a row.

\begin{figure}
\includegraphics[width=0.95\columnwidth]{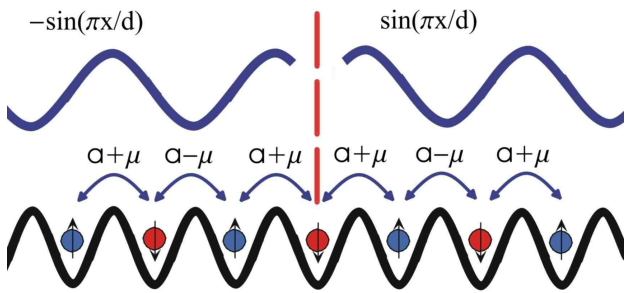}
\vspace{-3mm} \caption{Defect in the pattern of alternating hopping
amplitudes. The coefficient in the front of $\mu$ is determined by
the product $\varphi(x)\sin(\pi x/d)$ at the middle between the
lattice sites where the Wannier functions of the two lattice sites
overlap. The soliton, or a kink, profile of $\varphi(x)$ generates a
defect in the hopping amplitude pattern, so that the same coupling
matrix element appears twice. The defect is marked by a dashed (red)
vertical line. Here $\varphi(x)\sim x/|x|$.
     } \label{f3}
\end{figure}

For an even number of sites, or in the limit of an infinite lattice,
the two hopping amplitude profiles
\begin{equation}
\kappa_k=a \pm (-1)^k\mu\,,\label{dege}
\end{equation}
yield ground states with equal energies. We may join these two field
configurations at the center of the lattice by synthesizing a defect
in the alternation pattern in such a way that the left side of the
lattice represents the lower sign of the hopping term \eqref{dege}
and the right side of the lattice the upper sign; see Fig.~\ref{f3}.
The em field required to induce the desired coupling may be written
in the following form:
\begin{equation}
\Omega(\rv)= {\cal V}(\rv) + {\cal U}(\rv)\varphi(x) \sin\(
\frac{\pi x}{d}\) \,, \label{defectfield}
\end{equation}
where the field $\varphi(x)$, satisfying
$\varphi(h)=-\varphi(-h)=1$ for all $h\gg d$, exhibits a spatial
profile of a soliton, or a phase kink. Here $\varphi(x)$ has a phase
jump of $\pi$ at $x=0$, representing a topological phase
singularity. The corresponding hopping amplitude reads
\begin{equation}
\kappa_k = a + \varphi_k(-1)^k\mu\,,\label{defecthop}
\end{equation}
where $\varphi_k=\varphi(x_k)$; see Fig.~\ref{f3}.

This type of a coupling represents a superposition of a field with a
constant phase profile ${\cal V}$ (producing $a$) and the field
$\varphi(x)\sin (\pi x/d)$. The entire field configuration
\eqref{defectfield} could be prepared optically by synthesizing
multi-beam superposition states using diffractive optical
components. In particular, computer-generated holograms and spatial
light modulators can act as optical phase holograms to shape laser
fields in order to generate a sharp phase kink $\varphi(x)\sim
x/|x|$, combined with the sine wave and the field ${\cal V}$. The
holograms are typically calculated by specific algorithms that
optimize the phase and/or intensity profiles of the field according
to the merit function criteria. Optical holography has been
experimentally used to prepare phase singularities on light beams
\cite{leach04} and has been proposed as an efficient tool for phase
imprinting topological defects in atomic Bose-Einstein condensates
\cite{RUO05b,RUO01}.

\begin{figure}
\includegraphics[width=0.7\columnwidth]{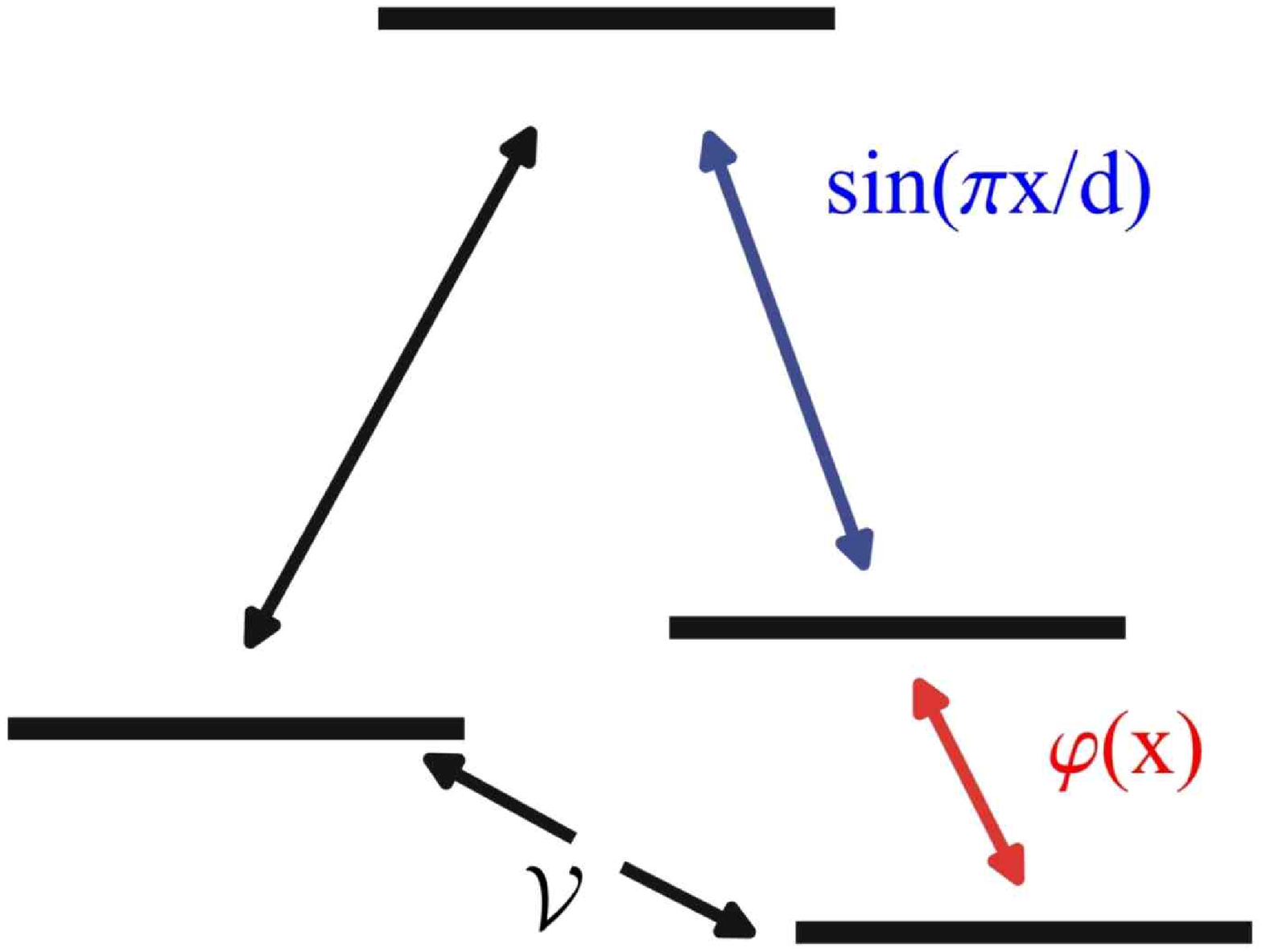}\vspace{3mm}
\includegraphics[width=0.67\columnwidth]{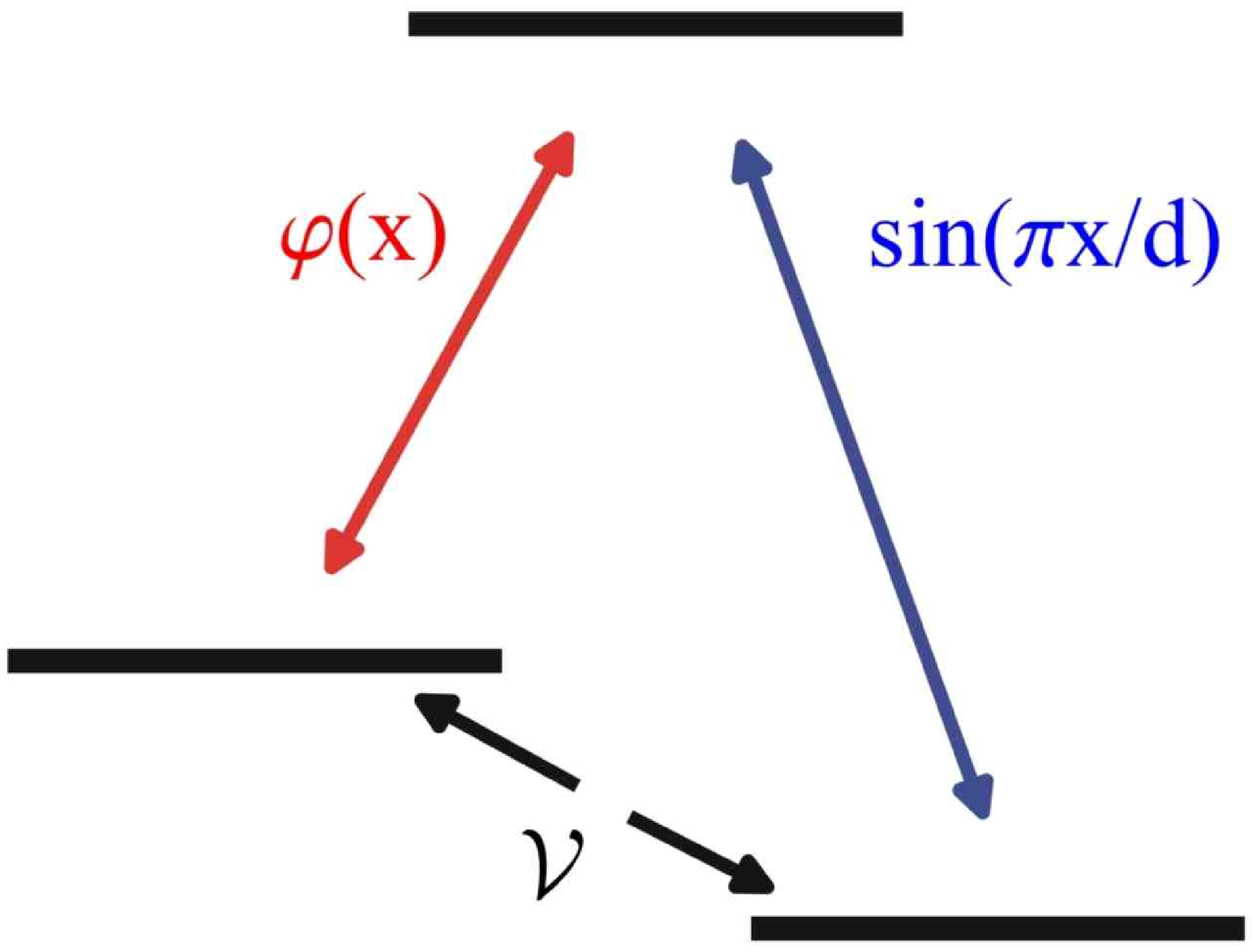}
\caption{Examples of the field configurations that are not based on
an optical hologram and that can generate the field amplitude in
\eq{defectfield}. Two different transition paths synthesize the
superposition between ${\cal V}$ and $\varphi(x) \sin(\pi x/d)$. The
field profile $\varphi(x) \sin(\pi x/d)$ is produced by two
successive transitions, so that the effective Rabi frequency is the
product of $\varphi(x)$ and $\sin(\pi x/d)$. On top, a far-detuned
microwave or rf field, generating $\varphi(x)$, is combined with an
optical Raman two-photon transition via a far-detuned electronically
excited intermediate level, where one of the transitions is driven
by the standing wave $\sin(\pi x/d)$. The effective Rabi frequency
for the corresponding three-photon transition is proportional to $\propto
\varphi(x)
\sin(\pi x/d)$. On bottom, an optical Raman two-photon transition
where one transition is driven by $\varphi(x)$ and the other one by
$\sin(\pi x/d)$.
     } \label{f4}
\end{figure}

Alternatively, one may generate the effective coupling field,
\eq{defectfield}, by superposing two different transition paths
between the states $|\up\>$ and $|\down\>$. One of the transition
paths, that may involve a multi-photon transition via intermediate
levels or a single-photon transition with no intermediate levels,
produces an effective Rabi frequency with a uniform phase profile,
representing the field ${\cal V}$ in \eq{defectfield} and $a$ in
\eq{defecthop}. The other transition path generates the field
$\varphi(x)\sin (\pi x/d)$. This could be obtained, e.g., with a
two-photon transition via an intermediate level, so that the first
transition is driven by $\varphi(x)$ and this is followed by the
second one driven by $\sin (\pi x/d)$; see Fig.~\ref{f4}.

The effective $\varphi(x)$ field might be produced either by making
use of the (rf or microwave) transition between the spin states, or
by means of an optical transition other than the one used to produce
the $\sin (\pi x/d)$ standing wave. In our studies of
fractionalization, the precise form of the field $\varphi(x)$ is not
very crucial, as the relevant physics depends on its asymptotic,
topological behavior. A phase profile with topological properties
similar to $\varphi(x)$ could be prepared, e.g., by means of a
standing em wave $\sin(q x)$ satisfying $q \ll \pi/ d$. The period
of $\sin(qx)$ needs to be chosen long so that along the lattice
there is only one node point located near the center of the
lattice. Such a 1D standing wave could be obtained using microwaves
or, alternatively, optical fields if the intersection angle between
the two laser beams could be chosen such that $\sin(\theta/2)\ll
\lambda/2d$, where $\lambda$ is the laser wavelength. In
Ref.~\cite{RUO02} we proposed also other potential experimental
schemes to prepare the hopping field based on em field couplings.

\subsubsection{Alternative scheme based on field gradients}

The scheme introduced in Ref.~\cite{RUO02} to prepare a nonuniform
hopping amplitude pattern for the atoms in an optical lattice has an
alternative proposed realization by Jaksch and Zoller \cite{JAK03}.
In Ref.~\cite{JAK03} a spatially alternating pattern of hopping
amplitudes was generated in a 2D lattice along one spatial direction
in order to induce a nonvanishing phase on atoms moving along a
closed path in the lattice, therefore simulating the effect of a magnetic
field on electrons.
This approach uses field gradients combined with an em-induced coupling.
In the following we adapt the basic scheme of Ref.~\cite{JAK03} to our
fractionalization study.

\begin{figure}
\includegraphics[width=0.7\columnwidth]{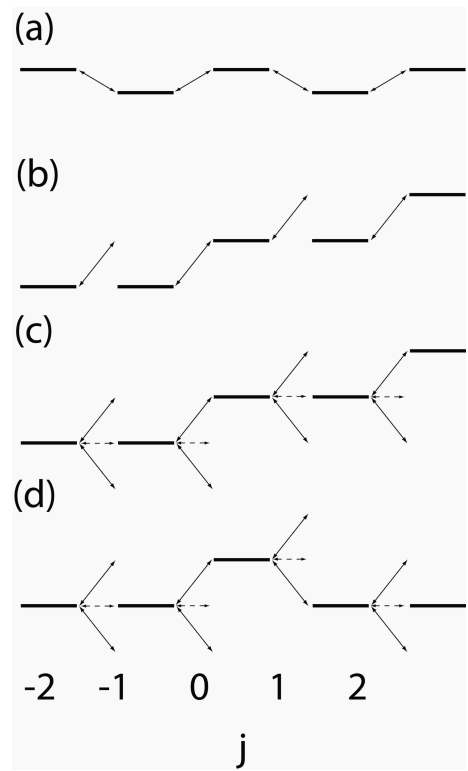}
\caption{A scheme for a dimerized lattice based on the interplay between
internal and center-of-mass energies of the atoms. (a) The
energies of the relevant internal states ($\up$ and $\down$) of an atom
at lattice sites
$j=-2,\ldots,2$ are marked by horizontal bars, with resonant
transitions between the states displayed by solid arrows. (b) After a
constant energy difference between adjacent lattice sites is added to the
atoms, for a given set of driving fields only half of the transitions can
be on resonance. (c) Another set of lasers frequencies is added, which
render also the remaining transitions resonant (dashed arrows). A
dimerized lattice has been prepared. Several
transitions without a resonant final state are are also indicated. (d) In
a variation of the scheme in part~(b), a constant energy difference is
added between every state pair to the left of $j=0$, and the negative of
the same energy difference between state pairs to the right of $j=0$. The
same coupling as in part (c) now realize a dimerized lattice with a
defect.
   } \label{zs}
\end{figure}

As in the previous section, we assume that the hopping of the atoms
between adjacent lattice sites is driven by em fields. To be specific,
let us think of two-photon transitions in the Raman configuration  with a
far-off resonant intermediate state. This setup is functionally
equivalent to transitions in an effective two level system with the
transition frequency $\omega$ equal to the difference of the energies of
the states
$\up$ and $\down$, driven by a field whose effective frequency is the
frequency difference between the two laser frequencies.  In
Fig.~\ref{zs}(a) we draw the scheme of transitions from site to
site. The horizontal lines represent the energies of the levels
$\up$ and
$\down$, alternating between adjacent sites, and resonant transitions
between the levels at the effective frequency of the driving
field $\omega$  are marked with arrows. In this basic lattice the
site-to-site couplings are the same.

Next assume that in some manner a potential has been
added to the atoms that grows linearly along the lattice.
Correspondingly, the Hamiltonian picks up an additional term
\begin{equation}
H_s = \xi\sum_{j=-N_h}^{N_h} jc_j^\dagger c_j\,.\label{static1}
\end{equation}
For notational simplicity the lattice sites
are numbered from $-N_h$ to $N_h$.
We specifically pick $\xi=\hbar\omega$, and imagine tuning the fields
driving the transitions to the effective frequency $2\omega$. As
demonstrated in Fig.~\ref{zs}(b), in this way every second transition is
on resonance. On the other hand, every second transition now has
the effective resonance frequency 0, and the field with the frequency
$2\omega$ assumedly becomes ineffective at transferring atoms in such
transitions. To compensate, let us add light fields to
make the effective frequency $0$ to drive the transition between the sites
that the potential~\eq{static1} decoupled, as in Fig.~\ref{zs}(c).
Resonant couplings between all adjacent sites have now been established.
To prepare for the final step of the argument, in Fig.~\ref{zs}(c) we have
also indicated a few possible transitions without a resonant
final level. As we denote symbolically by using  solid-line and
dashed-line arrows for the driving fields, the coupling coefficients for
the alternating transitions need not be equal. At this stage we have a
dimerized lattice.

In the last step we modify the added
potential to make the corresponding force change sign at the center of
the lattice, so that
\begin{equation}
H_s = \xi\sum_{j=-N_h}^{N_h} |j| c_j^\dagger c_j\label{static2}
\end{equation}
replaces~\eq{static1}.
Figure.~\ref{zs}(d) shows both the level scheme for the newly adjusted
lattice potential, and the possible transitions. This time around
two solid-arrow transition in a row are on resonance at the center, so
that we have established a defect in the couplings precisely as needed to
prepare a half-fermion.

The technical issue with these ideas is to set up the piecewise
linear potentials. A potential of the type~\eq{static1} could be
generated, say, with a magnetic field ramp that shifts Zeeman
levels, using ac stark shifts due to added off-resonant laser or
microwave fields, in the case of polar molecules static electric fields,
or even by accelerating the optical lattice. The reversal of the
force at the center of the lattice in~\eq{static2} could, e.g., in
the case of off-resonant lasers be prepared using optical holograms
or spatial light modulators. The required field gradients may also
be generated by shining a laser through an absorption plate with a
linearly changing optical thickness.

\subsubsection{Remarks}

We have demonstrated how em field-induced hopping of atoms between
adjacent lattice sites can be a very useful tool in engineering a
rich variety of atomic lattice Hamiltonians. A nonuniform field
amplitude, with a periodic spatial profile, is sufficient to prepare
a dimerized lattice. One may also introduce a defect in the regular
pattern of hopping amplitudes. We will next show that the hopping
amplitude profile with such a defect in the pattern of alternating
coupling matrix elements results in a bound state and a fractional
fermion number, localized around the defect. Such a coupling
frequency also converts the atomic lattice Hamiltonian (\ref{ham})
in the continuum limit into the Dirac Hamiltonian exhibiting
fractional particle number in relativistic field theory.

\subsection{Fractional fermion in a finite lattice}\label{finitesec}

In this Section we consider the Hamiltonian
\eqref{ham} in the presence of the hopping field with a defect,
\eq{defecthop}, along the lines of Ref.~\cite{BEL83}, for a finite
lattice and without a continuum limit. We use hard-wall boundary
conditions, i.e., the lattice simply terminates at the edges as if the
first sites past the ones included in the lattice were forced to be
empty. We demonstrate how a lump with the particle number expectation
value of a half emerges on top of the a homogeneous background fermion
density. We also discuss numerical techniques used to study the
fractional fermion, both as it comes to the expectation value and the
fluctuations of the atom number, and present representative examples.

\subsubsection{Elementary half-fermion}

Suppose we can find fermion operators $\gamma_n$ of the form
\begin{equation}
\gamma_p = \sum_k U_{kp} c_k = \sum_k a_k c_k \label{GDF}
\end{equation}
such that the Hamiltonian~(\ref{ham}) becomes
\begin{equation}
\frac{H}{\hbar}= \sum_p \omega_p \gamma^\dagger_p \gamma_p\,,
\label{DIA}
\end{equation}
then we have diagonalized the Hamiltonian. Any one of the operators
$\gamma_p$, call it $\gamma$, should then  satisfy
\begin{equation}
\left[\gamma, \frac{H}{\hbar}\right] = \omega\gamma. \label{EQM}
\end{equation}
It follows readily from Eqs.~(\ref{GDF})--(\ref{EQM}) that the
column vectors $a_k$ of the matrix $U_{kp}$ are the orthonormal
eigenvectors of the eigenvalue problem
\begin{equation}
(\epsilon_k -\omega)a_k -  \kappa_k a_{k+1} - \kappa_{k-1} a_{k-1} =
0\,, \label{EVP}
\end{equation}
which also yields the corresponding eigenvalues $\omega_p$.
As in~\eq{EVP}, the couplings $\kappa_k$ may be assumed
real without a loss of generality, and so we take the matrix $U$ to be
real and orthogonal
\begin{equation}
U_{kp}^{-1}=U_{pk}\,.\label{orthogo}
\end{equation}

We consider a sharp defect $\varphi(x)\sim x/|x|$ in the hopping
amplitude profile. As in Ref.~\cite{BEL83}, we take the number of
lattice sites to be $N_s\equiv 2N_h+1\equiv 4n+1$, where $n$ itself
is an integer, and number the sites with integers ranging from
$-N_h$ to $N_h$. For illustration, pick $n=2$, use an {\rm x} to
denote a lattice site, and $\pm$ the couplings $a\pm\mu$, then our
lattice with the couplings reads
\begin{equation} {\rm x}-{\rm x}+{\rm x}-{\rm x}+{\rm x}+{\rm x}-{\rm
x}+{\rm x}-{\rm x}\,
\end{equation}
The defect in the middle, two consecutive plus signs, is the crux of
the matter.

We assume here a uniform lattice with the effective energy splitting
between the adjacent sites $\delta=0$, so that $\epsilon_k\equiv0$.
It is easy to see from the structure of \eq{EVP} that if $\omega$ is
an eigenvalue, then so is $-\omega$ (the conjugation
symmetry); and the eigenvectors transform into one another by
inverting the sign of every second component. We will label the
eigenvectors as $-N_h\ldots N_h$ in ascending order of frequency,
and assign the labels $\pm p$ to such $\pm$ pair of states.
Correspondingly, the transformation matrix $U$ satisfies
\begin{equation}
|U_{kp}| = |U_{k,-p}|\,.
\end{equation}

But under our assumptions, the number of eigenvalues and eigenstates
is odd. The $\pm$ symmetry implies that an odd number of the
eigenvalues must equal zero. Except for special values of the
couplings $a$ and $\mu$, there is one zero eigenvalue. We call the
corresponding eigenstate the zero state. Provided $a$ and $\mu$ have
the opposite signs and $|a|>|\mu|$, all odd components in the zero
state equal zero and the even components are of the form
\cite{signcomment}
\begin{equation}
x_k =
x_0\left({-\displaystyle\frac{a+\mu}{a-\mu}}\right)^{|k|/2}\,.\label{zerosol}
\end{equation}
The normalizable zero state becomes  the narrower, the closer in
absolute value $a$ and $\mu$ are. For a broadly distributed bound
mode, $|a|\gg |\mu|$, we obtain from \eq{zerosol} the limit
\begin{equation}
\frac{x_k}{ x_0} \simeq (-1)^{|k|/2}
\exp{\(-|x_k-x_0|/\xi\)}\,,\label{zerosolco}
\end{equation}
where $\xi$ is the atomic correlation length calculated in
\eq{correlation}. Since the size of the zero state depends on the
relative strength of the superposed electromagnetic fields via the
correlation length, it could be varied experimentally. This is
different from the polymer case, where the size of the bound state
is fixed.

Suppose next that the system is at zero temperature, and contains
$N_f=N_h+1$ fermions. The exact eigenstates $p$ at the ground state
are then filled up to zero state and empty at higher energies, with
occupation numbers $n_p$= 0 or 1, where
\begin{equation}
n_p\equiv \<\gamma_p^\dagger\gamma_p\>\,.
\end{equation}
The number operator for the fermions at site $k$ correspondingly
reads from Eqs.~\eqref{GDF} and~\eqref{orthogo}
\begin{equation}
c^\dagger_k c_k=\sum_{pq} U_{kp} U_{kq}\gamma^\dagger_p \gamma_q\,,
\end{equation}
so that the expectation value of the fermion number at the site $k$ is
\begin{align}
\langle c^\dagger_k c_k \rangle  =& \sum_{p=-N_h}^0 |U_{kp}|^2
= \sum_{p=-N_h}^{-1}|U_{kp}|^2 + |U_{k0}|^2\nonumber\\
=& \half\sum_{p=-N_h}^{-1}|U_{kp}|^2
+\half\sum_{p=+1}^{N_h}|U_{kp}|^2+ |U_{k0}|^2\,. \label{expnumber1}
\end{align}
Here we have explicitly separated the zero state and used the
symmetry $|U_{kp}| = |U_{k,-p}|$. The matrix $U_{kp}$ defines a
complete basis and we can use the corresponding orthogonality of $U$
by combining the first two terms and one-half of the third term on
the last line of \eq{expnumber1}:
\begin{equation}
\langle c^\dagger_k c_k \rangle  =
\half\sum_{p=-N_h}^{N_h}|U_{kp}|^2 + \half|U_{k0}|^2 = \half +
\half|U_{k0}|^2\,. \label{expnumber}
\end{equation}
By virtue of the same orthogonality, localized with the zero state
there is a lump with $\half\sum_{k} |U_{k0}|^2=\half$ fermions on
top of a uniform background of half a fermion per site.

This lump is the celebrated half of a fermion
\cite{JAC76,NIE86,jackiw}. It is the
result of adding a fermion to the zero state which has a nonvanishing
occupation probability only for every second site. Therefore the
half-fermion rising from the constant background resides on every second
lattice site, here on the even-numbered sites. For the remaining lattice
sites the mean occupation number is exactly a half.

As a result of the defect the normalizable zero state is localized around
the phase kink. This zero state creates a fractional deficit of states in
the valence and the conduction bands, as can be observed in
\eq{expnumber1}. In the presence of the conjugation symmetry, the
density of states is a symmetric function of the energy, and both
bands have locally a deficit of one-half of a state.

\subsubsection{Fermion numbers and their fluctuations}\label{FLUSTU}

To demonstrate the precise meaning of the fractional fermion, we
define two smoothed fermion number operators using an envelope function
$\alpha_k$ that covers the zero state around the phase kink;
\begin{equation}
\h N  = \sum_k \alpha_k c_k^\dagger c_k, \quad
\tilde{N}= \sum_k \alpha_k \( c_k^\dagger c_k -\half
\)\,.\label{fracfermion}
\end{equation}
The idea is that an experiment with a limited resolution may be
expected to address many lattice sites at once. Alhtough the numbers
$\alpha_k$ are real in all of our worked-out examples, for maximum
flexibility we allow
them to be complex. The difference between the operators $\hat{N}$
and
$\tilde{N}$ is that in the latter we have subtracted the constant
background fermion number
$\half$ at each lattice site.

Let us so far assume that the
numbers $\alpha_k$ vary little from site to site, and that their values in
a region that covers the half-fermion lump is well approximated by unity.
We then have from
\eq{expnumber}
\begin{equation}
\<\tilde N\> = \half \sum_k \alpha_k |U_{k0}|^2=\half\,,
\end{equation}
which just reiterates the observation that the lump contains, on the
average, half of a fermion above the uniform background.

There is more to a fractional fermion number, however, than the
average density. So far we have only dealt with the {\it
expectation\/} values of the atom numbers. However, the {\it
fluctuations\/} can be small as well, indicating that the half-fermion
represents
an eigenstate of the operator
$\tilde N$~\cite{BEL83,kivelson}. Since the
fluctuations are the same for the operators $\hat N$ and $\tilde N$,
we focus on the former. The expectation value of $\hat N$ is
\begin{eqnarray}
\langle\hat N \rangle &=& \sum_k \alpha_k \langle c^\dagger_k c_k \rangle
\nonumber\\
&=& \sum_k \alpha_k U_{kp}U_{kq}\langle\gamma_p^\dagger\gamma_q
\rangle\nonumber\\
&=& \sum_k \alpha_k U_{kp}^2 n_p\,,
\label{CSUM}
\end{eqnarray}
and for the square we have
\begin{eqnarray}
\lefteqn{\langle {\hat N}^\dagger {\hat N} \rangle}\nonumber\\
    &=& \sum_{kl} \alpha^*_k
\alpha_l \langle c^\dagger_k c_k c^\dagger_l c_l\rangle\nonumber\\
&=& \sum_{klpqrs} \alpha^*_k \alpha_l U_{kp}U_{kq}U_{lr}U_{ls}
\langle \gamma^\dagger_p \gamma_q \gamma^\dagger_r
\gamma_s\rangle\nonumber\\ &=& \sum_{klpq} \alpha^*_k \alpha_l \left[
U^2_{kp}U^2_{lq} n_p n_q +U_{kp}U_{kq}U_{lp}U_{lq}(n_p-n_pn_q)
\right]\nonumber\\ &=& |\langle\hat N\rangle|^2 + (\Delta N)^2\,,
\label{FRS}
\end{eqnarray}
where the part characterizing the fluctuations is
\begin{align}
(\Delta N)^2 = & \langle{\hat N}^\dagger{\hat N}\rangle - |\langle {\hat
N}\rangle|^2\nonumber\\= &\sum_{klpq} \alpha^*_k \alpha_l
U_{kp}U_{kq}U_{lp}U_{lq}n_p(1-n_q)\,. \label{FFLU}
\end{align}
Although we always use the zero-temperature Fermi sea with the
occupation numbers $n_p$ = 0 or 1 for the $\gamma_p$ fermions in our
explicit examples, this assumption has not been used in the analysis
of the fluctuations. With the appropriate thermal occupation numbers
$n_p$, expressions such as \eq{FFLU} apply also at finite
temperature \cite{AD,DR}.

\begin{figure}
\includegraphics[width=7.5cm]{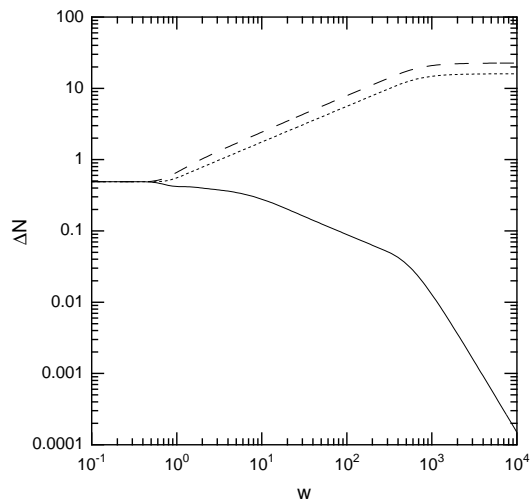}
\vspace{-10pt} \caption{Coherent atom number fluctuations $\Delta N$
under a Gaussian (solid line) and alternating-sign Gaussian (dashed
line) envelopes, $\alpha_k^G$ and $\alpha_k^A$, and the
corresponding incoherent atom number fluctuations $(\Delta N)_i$
(dotted line) as a function of the width of the Gaussian $w$. The
lattice parameters are $N_s=1025$, $N_f=513$, and $\mu=-0.1\,a$. }
\label{FLUCTS}
\end{figure}

For illustration we use the Gaussian shape,
\begin{equation}
\alpha_k^G = e^{-(k/w)^2}\,.\label{gaussprof}
\end{equation}
We take a lattice with $N_s=1025$ sites, pick the parameters
$\mu=-0.1\,a$, put in $N_f=513$ fermions so that the zero state is
the last filled stated, and find the rms fluctuations  of the
fermion number $\Delta N$ as a function of the width of the weight
function $w$. The result is shown as the solid line on a log-log
plot in Fig.~\ref{FLUCTS}. The notch around $w=1$ indicates that at
this point the weight factors $\alpha_k$ start to span several
lattice sites. Another break in the curve is seen at about $w=10$
when the weight function covers the whole zero state. Thereafter the
fluctuations behave as $\Delta N\propto w^{-1/2}$.  The fermion
number $\hat N$ under the weight function becomes more sharply
defined as the region for averaging grows broader. Finally, at
$w\sim500$, the weights $\alpha_k$ effectively cover the entire
lattice. The fluctuations then decrease even faster with increasing
$w$, as is appropriate for the fixed fermion number in the lattice
as a whole.

Let us next suppose that the weights
$\alpha_k$ can be arranged to alternate in sign
from site to site. Specifically, we use
\begin{equation}
\alpha_k^A = (-1)^k e^{-(k/w)^2}\,.\label{agaussprof}
\end{equation}
Since the half-fermion is confined entirely to the
even-numbered sites, this weight has the effect that in the limit
$w\rightarrow\infty$ the constant background is automatically subtracted
as far as mean atom numbers go;
$\langle
\hat N\rangle =
\langle
\tilde N\rangle =
\half$. Regarding fluctuations, we again use~\eq{FFLU}. The resulting
$\Delta N$ is plotted in Fig.~\ref{FLUCTS} as a function of
the width $w$ as the dashed line for the same problem parameters that
were used for the single-sign Gaussian $\alpha_k^G$. Once the envelope
covers several sites, the fluctuations for the models
$\alpha_k^G$ and $\alpha_k^A$ go separate ways, and in the latter case
actually increase with increasing width $w$ until the whole lattice is
covered.

Our final case of fluctuations is the case of simply adding the
particle number fluctuation at each site with the weights $|\alpha_k|^2$,
as if the fluctuations were independent. We thus have the
``incoherent'' fluctuations,
\begin{align}
(\Delta N)^2_i =& \sum_{k} |\alpha_k|^2 \left(\langle c^\dagger_k
c_k \rangle - \langle c^\dagger_k c_k \rangle \langle c^\dagger_k
c_k\rangle\right)
\nonumber\\
=&\sum_{kpq} |\alpha_k|^2 U_{kp}^2U_{kq}^2n_p(1-n_q)\,.
\label{incfluc}
\end{align}
This is what becomes of the ``coherent'' fluctuations of~\eq{FFLU} if only
the diagonal terms with $k=l$ are kept. A physical motivation for such a
rule could be that the coefficients
$\alpha_k$ are uncorrelated random numbers (with zero mean).

The incoherent fluctuations
$(\Delta N)_i$ are plotted in Fig.~\ref{FLUCTS} as a dotted line along
with the coherent fluctuations for the two envelopes $\alpha_k^G$
and $\alpha_k^A$; which envelope is used in the incoherent case does not
matter. The magnitude of the incoherent fluctuations is between the
magnitudes of the fluctuations for the two coherent cases. As soon as the
widths of the Gaussians are sufficient to cover several lattice sites, the
coherent fluctuations for the alternating-sing Gaussian $\alpha_k^A$ are
about a factor of $\sqrt{2}$ times the incoherent fluctuations.

In the standard half-integer fermion number
argument one uses a smooth envelope such as $\alpha_k^G$ and subtracts
a non-fluctuating neutralizing background of $\half$ charge per
lattice site, whereupon $\langle \tilde{N}\rangle
\rightarrow\half$ and
$\Delta N\rightarrow 0$ with an increasing envelope width $w$. The
intermediate regime that occurs once the zero state is covered is
the crux of the matter. Not only does the expectation value of
fermion number $\tilde N$ equal $\half$, but the fluctuations are
small and $\tilde N$ therefore has an eigenvalue \half.

A perusal of the dotted line for the incoherent fluctuations in
Fig.~\ref{FLUCTS} qualitatively suggests that the fluctuations of
the fermion number at each individual site are comparable to what one
would expect if each site simply had the fermion number of either 0 or 1
with the equal probabilities of \hbox{$1\over2$}. On the other hand, the
atom number fluctuations appear to be anticorrelated between
adjacent sites. Thus, if one sums up the atom numbers with a weight that
has a sign alternating from site to site (dashed line), the resulting
fluctuations in the summed atom number are larger than they would be
for uncorrelated fluctuations. Also, if one sums over the atom numbers
with a weight that varies little from site to site (solid line),
such anticorrelated fluctuations tend to cancel in the sum. In short,
anticorrelations of atom number fluctuations  between neighboring sites
undoubtedly contribute to the impression of a sharp eigenvalue for a
smoothly weighted sum of occupation number operators. The role of
longer-range correlations (next-nearest sites, etc.) deserves further
study, but it will not be undertaken here.

\subsubsection{Variations of the half-fermion}

The main point to emphasize about the derivation of~\eq{expnumber}
is that the fractional particle number is the property of the
system as a whole. Adding the last fermion that occupies the zero state
adds one particle, but makes a soliton containing half of a
fermion. This means that before the addition there was a hole in the
density distribution with half of a fermion missing, an observation
easily verified by a calculation similar to the one that lead
to~\eq{expnumber}. The hole clearly depends on the
shapes of the wave functions and on the occupation numbers of the
non-zero states, hence so does the half-integer fermion number.

Fractionalization is a robust phenomenon. Something akin to a
localized zero state occurs as soon
as the regular alternation of the couplings between adjacent states
gets out of rhythm around a defect. In particular, the defect does
not have to be confined to one lattice site; this should make the
experiments easier. Moreover, a spike with about half a fermion
standing above the background may be seen when the particle number
deviates  from $(N_s+1)/2$ by, say, ten per cent; and a hole of
about half a fermion may similarly persist even if the particle
number is well under $(N_s+1)/2$.

The total number of atoms, of course, must remain an integer, and the
optical lattice has a finite size. In our examples with $N_s=4n+1$ sites
and $N_f=2n+1$ fermions, the background occupation of \half\ makes $2n+\half$
fermions, and the half of the fermion in the peak adds up to make the
correct total fermion number $2n+1$. If the number of sites is even,
the system
compensates by putting another soliton at the edge of the lattice,
similar to Shockley edge states \cite{SHO39}. If the
number of sites is even and the number of fermions precisely half of it,
numerically, all solitons may vanish.
The zero state is then doubly degenerate, and, as
we have not tried to assert any control over the state in the degenerate
subspace that gets the last fermion, the numerics decides for us.

The restriction $|a|>|\mu|$ and $a$ and $\mu$ exhibiting the
opposite signs could be relaxed to read $|a+\mu|<|a-\mu|$. If this
condition is violated, the primary soliton, the counterpart of the
half-fermion in our examples, emerges at the edge of the lattice,
and the possible extra soliton at the center.


\section{Fermion number 1/3}\label{onethreesec}

We may also generalize the half-fermion concept to fractionalization
to values $1/n$, for any integer $n$, by considering defects in an
$n$-fold degenerate ground state, analogously to the generalization
of the polymer system \cite{HEE88,SCH85}. To illustrate, we consider
the fractional fermion number $1/3$. We consider the ground state
with a $1/3$ atomic filling factor. We assume that the hopping
amplitude between the lattice sites reads
\begin{equation}
\kappa_k^{(p)}= a + A_k^{(p)}\mu \,.\label{3kappa}
\end{equation}
Here the phase factors $A_k^{(p)}$ exhibit a periodicity of three
lattice sites and define a three-fold degenerate ground state (in
the limit of an infinite lattice) for the values $p=0,1,2$. We
choose
\begin{equation}
A_k^{(p)} = {1\over3} -{4\over3} \cos{\[{4\pi (k+p)\over
3}\]}\,.\label{threefold}
\end{equation}
Then for $p=0$ we have
\begin{equation}
{\rm x}+{\rm x}+{\rm x}-{\rm x}+{\rm x}+{\rm x}-{\rm x}+{\rm x}+{\rm
x}-{\rm x}\cdots\,,
\end{equation}
where we again use an {\rm x} to denote a lattice site and $\pm$ the
couplings $a\pm\mu$. Similarly, for $p=1$ we obtain
\begin{equation}
{\rm x}+{\rm x}-{\rm x}+{\rm x}+{\rm x}-{\rm x}+{\rm x}+{\rm x}-{\rm
x}+{\rm x}\cdots\,.
\end{equation}
\begin{figure}
\includegraphics[width=0.9\columnwidth]{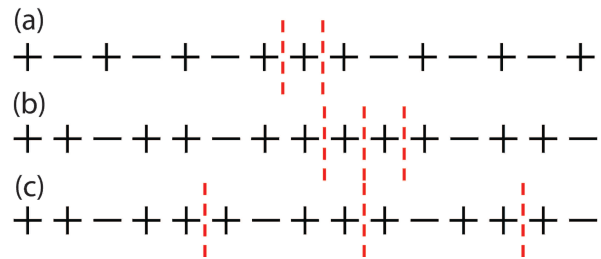}
\vspace{-3mm} \caption{The nonuniform spatial profile for the
hopping amplitudes. (a) Here `$\pm$' denote the couplings $a\pm\mu$
in \eq{ome2} where one of the signs of the hopping amplitudes has
been changed. Introducing the sign switch generates two defects that
are marked by the vertical lines. (b) Here `$\pm$' denote the
couplings $a\pm\mu$ in \eq{3kappa}, with $A_k^{(p)}$ given by
\eq{threefold}, where one of the signs in the middle has been
switched. In this case the sign change introduces three defects
(marked by vertical lines). The first defect represents the boundary
between the $p=0$ and $p=2$ three-fold degenerate configurations of
\eq{threefold}; the second between $p=2$ and $p=1$; and the third between
$p=1$ and $p=0$. Part (c) represents the same configuration as shown in
(b), but with the three defects moved apart.
     } \label{onethree}
\end{figure}

In the dimerized lattice system with the coupling \eqref{ome2}, one
may easily observe that changing the sign of one of the hopping
amplitudes corresponds to adding {\it two} identical defects in the
systems [Fig.~\ref{onethree}(a)] and by changing the number of
atomic states in the vacuum by one (for the ground state of the
average of 1/2 atoms per site). The half fermion results from
sharing the extra state evenly between the defects. Similarly, one
may show that changing one of the signs of the hopping amplitudes in
\eq{3kappa} corresponds to adding one atomic state in the $1/3$
filled vacuum. It is relatively easy to see that such a sign switch
can be viewed as {\it three} identical defects that join together
the three degenerate vacuum configurations of \eq{threefold},
representing the boundaries between the $p=0$ and $p=2$, $p=2$ and
$p=1$, as well as the $p=1$ and $p=0$ states; see
Fig.~\ref{onethree}(b-c). If the defects are moved wide apart
compared to the size of the kink bound states, the extra atomic
state has to be split equally between the defects, resulting in the
$1/3$ fermion number.

\section{Low energy continuum limit and fractionalization in
relativistic field theory}\label{relsec}

In the continuum limit the Hamiltonian (\ref{ham}) [with
$\Omega(\rv)$ defined in Eqs.~{(\ref{defectfield})}
and~(\ref{defecthop})] can be transformed \cite{HK78,TLM80,JAC83}
into a relativistic Dirac Hamiltonian exhibiting fractional fermion
number in quantum field theory
\cite{JAC76}. The transformation is mathematically the same as in the
1D polymer case
\cite{HK78,TLM80,JAC83,NIE86,HEE88}, but the physics is rather
different: in our
atomic system the kink $\varphi(x)$ appears in the em-induced
coupling between the internal atomic states, and not as a physical
domain wall kink.

A particular advantage of the continuum quantum field theory is that it is
amenable to a simple description. The continuum limit of the
low-energy expansion corresponds to the linearization of the band
structure about the Fermi surface by writing
$\epsilon(p)\simeq\epsilon(p_F)+(p-p_F)\epsilon'(p_F)$, with the
Fermi momentum $p_F=\pi/2d$, and keeping only terms to leading order
in small $d$. It becomes accurate in the dilute gas limit, when the
atomic correlation length $\xi\simeq |ad/\mu|$, from
\eq{correlation}, is much larger than the lattice spacing. In the
continuum limit we write the fermionic annihilation operators for
the even and odd sites as continuous functions of the lattice
spacing $d$:
\begin{align}
c_{2k+1}&\equiv(-1)^{(2k+1)/2}\sqrt{2d}\,u[(2k+1) d],\\
c_{2k}&\equiv(-1)^{k}\sqrt{2d}\,v(2k d)\,.
\end{align}
With these identifications, the continuum limit proceeds exactly as
in the polymer case
\cite{NIE86}. To leading order in small $d$ we obtain
\begin{align}
u'(2 n d) &\simeq \big[ u(2 n d+d)-u(2n d-d)\big]/2d,\\  v'(2n d+d)
&\simeq \big[v(2n d+2d)-v(2n d)\big]/2d,\\  u(2n d) &\simeq
\big[u(2n d+d)+u(2n d-d)\big]/2,\\ v(2n d+d) &\simeq \big[ v(2n
d+2d)+v(2n d)\big]/2\,.
\end{align}
Here $u'(n d)$ denotes a discrete spatial derivative of $u$. By
setting $\epsilon_k=(-1)^k \delta/2$ and with the hopping determined
by \eq{defectfield}, the Hamiltonian \eqref{ham} reads
\begin{align}
H/\hbar &= -2i d^2 a \sum_n [u^\dagger(n d)v'(n d) + v^\dagger(n
d)u'(n d)]\nonumber\\ &+ \frac{\delta d}{2}\sum_n [u^\dagger(n d) u(n
d)-v^\dagger(n d) v(n d)]\nonumber\\ &+2i\mu d \sum_n\varphi(n d)
[u^\dagger (n d) v(n d)- v^\dagger (n d) u(n d)]\,. \label{hamcon}
\end{align}
In the continuum limit we replace $n d\rightarrow x$ and
$d\sum_n\rightarrow\int dx$.  Combining the two states into a
two-component spinor
\begin{equation}
\Psi(x)\equiv
\begin{pmatrix}
u \\
v
\end{pmatrix}
\end{equation}
and making the transformation: $\Psi\rightarrow
\exp{(i\pi\sigma^3/4)}\Psi$, we may express Eq.~(\ref{hamcon}) as
the relativistic (1+1)-dimensional Dirac Hamiltonian
\cite{JAC76,NIE86} for the spinor $\Psi(x)$ coupled to a
bosonic condensate $\varphi(x)$,
\begin{equation}
H=\int dx\, [c\hbar\Psi^\dagger\sigma^2\frac{d\Psi}{i dx}+\hbar
g\varphi\Psi^\dagger\sigma^1\Psi+mc^2 \Psi^\dagger\sigma^3\Psi]\, .
\label{dir}
\end{equation}
Here we have identified $c=2da$, $g=2\mu$, and $m=\hbar\delta/(8d^2
a^2)$. The $\sigma^i$ denote the Pauli spin matrices, $g$ the
coupling coefficient, $m$ the fermionic mass, and $c$ the speed of
light. In the atomic system the spinor components refer to the two
internal atomic levels. The corresponding eigenvalue system reads
\begin{align}
\( c\hbar {d\over dx}+\hbar g \varphi \) u(x) - m c^2 v(x) =& E
v(x),\nonumber\\ \( -c\hbar {d\over dx}+\hbar g \varphi \) v(x) + m
c^2 u(x) =& E u(x)\,.\label{diraceqn}
\end{align}
Note that in the relativistic Dirac equation \eqref{diraceqn} for the
fermionic atoms both the speed of light and the coupling coefficient
may be controlled by the different em field amplitudes $a$ and
$\mu$. The mass of the Dirac fermions vanishes at the exact
resonance coupling between the two atomic levels and is linearly
proportional to the detuning that can be accurately changed in the
experiments. Moreover, in the relativistic theory, as in our atomic
scheme, the bosonic field $\varphi(x)$ can be taken to be a static
classical background field. Thus, the atomic Hamiltonian \eqref{ham}
with a defect in the pattern of alternating hopping amplitudes
[\eq{defecthop}] can also be written in the continuum limit in the
form (\ref{dir}) and~\eqref{diraceqn}.

Fractionalization in the relativistic theory is described  as
follows \cite{JAC76,NIE86}. Suppose the  bosonic field $\varphi$ in
\eq{dir} has a doubly-degenerate ground state, for example arising
from a double-well potential $V(\varphi)\propto
(\varphi^2-\gamma^2)^2$. There are two minima,
$\varphi(x)=\pm\gamma$, and the reflection symmetry
$\varphi\leftrightarrow -\varphi$ is spontaneously broken.
Consequently there are topological solitons $\varphi(x)$ (also known
as ``kinks'') interpolating between these two degenerate minima. For
weak coupling, we make a Born-Oppenheimer approximation, so that the
fermion number must now be defined through the second quantization
of the fermion field in the presence of this background soliton.
Being a Dirac particle, the fermion has both positive and negative
energy single-particle eigenstates, and therefore an associated
Dirac negative energy sea in which the negative energy states are
occupied. This Dirac sea is the ultimate physical origin of the
fractional fermion number \cite{jackiw}.

The second-quantized number operator can be defined in two
equivalent ways. The formal relativistic quantum field theory
definition is \cite{BJD}
\begin{equation} N\equiv \frac{1}{2}\int dx\, [\Psi^\dagger(x),\Psi(x)]\, ,
\label{ncomm}
\end{equation}
with the commutator ensuring that the fermion particle number
vanishes in the free vacuum. Equivalently, one defines the physical
particle number in the soliton sector as being measured {\it
relative to} the free vacuum sector, so that the fermion number
density is
\begin{eqnarray}
\rho(x)= \int_{-\infty}^{0^-} dE \left(|\Psi_E(x)|^2-|\psi_E(x)|^2\right)
\label{sub}
\end{eqnarray}
where $\Psi_E$ ($\psi_E$) are the  fermion
single-particle energy eigenstates in the soliton (free) vacuum
sector. Once again, by construction the fermion particle number
vanishes in the free vacuum.

It is easiest to exhibit the fermion number fractionalization when
$m=0$ in the Dirac Hamiltonian (\ref{dir}). In this case the
Hamiltonian has a conjugation symmetry that pairs positive and
negative energy states. For every eigenvalue $\epsilon$ of
\eq{diraceqn} there exists an eigenvalue $-\epsilon$, and the
corresponding eigenfunctions are paired according to
$\Psi_{-\epsilon}=\sigma^3\Psi_\epsilon^*$. However, the nontrivial
feature of this fermion-soliton system is that there is also
a zero-energy bound state $\Psi_0(x)$, and furthermore, this state
is localized at the soliton jump:
\begin{equation}
\Psi_0(x)= \begin{pmatrix}
A \exp{\[-{g\over c} \int^x_0 d y\, \varphi(y)\]}  \\
0
\end{pmatrix}\,.
\end{equation}
This state is self-conjugate, $\sigma^3 \Psi_0=\Psi_0$, and results
in a doubly-degenerate soliton sector vacuum \cite{JAC76}.

Thus, the mode expansion of the second quantized fermion field
operator has the form
\begin{eqnarray}
\Psi(x, t)&=&a\, \Psi_0(x)\label{second}\\
&&\hskip -1cm +\int dk\, \left(e^{-iE_k t} b_k \,\Psi_{E_k}(x)+
e^{iE_k t} c^\dagger_k \,\Psi^*_{E_k}(x) \right)
\nonumber
\end{eqnarray}
where $b_k$ and $c_k$ denote annihilation operators for continuum
fermion and antifermion modes (respectively), while the operator $a$
is the annihilation operator for the self-conjugate zero-energy
state. Then the fermion number operator (\ref{ncomm}) in the
presence of the soliton reads \cite{NIE86}:
\begin{align}
N& = \half [a^\dagger,a] + \half \int dk \, \( [b_k^\dagger,
b_k]+[c_k, c_k^\dagger]\) \nonumber\\
&=a^\dagger a-\half+\int dk \,(b^\dagger_k b_k-c^\dagger_k c_k)\,,
\label{ncomm2}
\end{align}
Note that the operators $a$ and $a^\dagger$ couple the two
degenerate fermion-soliton ground states in which the zero energy
single-particle state is either occupied or not. Thus, it follows
from (\ref{ncomm2}) that the ground-state fermion-soliton states
possess fractional fermion numbers $\pm1/2$.

The same conclusion is reached \cite{jackiw} using the subtracted
definition (\ref{sub}) of the number density operator, if we simply
combine the expression (\ref{sub}) with the completeness of the two
sets of eigenstates, $\{\Psi_E\}$ and $\{\psi_E\}$, in the soliton
and vacuum sector, respectively:
\begin{eqnarray}
&&\int_{-\infty}^{0^-} dE\,
|\Psi_E(x)|^2+|\Psi_0(x)|^2+\int_{0^+}^{+\infty}dE\,
|\Psi_E(x)|^2\nonumber\\
&&\hskip 2cm =\int_{-\infty}^{+\infty}dE\, |\psi_E(x)|^2
\label{completeness}
\end{eqnarray}
Thus, in the fermion-soliton vacuum defined with only negative
energy states filled (i.e., with the zero mode empty), the fermion
number density (\ref{sub})  is
\begin{eqnarray}
\rho(x)=-\frac{1}{2} |\Psi_0(x)|^2
\label{rho0}
\end{eqnarray}
which yields fermion number $-1/2$. Similarly, if the zero energy
state is filled, the fermion number is $+1/2$, consistent with the
result above from (\ref{ncomm2}).

In fact, it is not just the expectation value of the fermion number
that is half-integer; with a suitable physical definition of the
number density operator as a smeared operator defined over a
physical sampling region, it can also be shown that these
half-integer values are {\it eigenvalues} of the number operator
\cite{kivelson,BEL83}. The fermion-soliton system yields {\it
eigenstates} of this physical number operator. Furthermore, the
fractional part of the fermion number  has a topological character:
it is insensitive to local deformations of the bosonic soliton
field, depending only on its asymptotic behavior \cite{NIE86}.

For $m\neq0$, the situation becomes even more interesting \cite{GW}.
In this case, the conjugation symmetry of the Dirac Hamiltonian
(\ref{dir}) is broken and the positive and negative energy states
are no longer paired in a simple way. In the limit of a slowly
varying soliton (on the scale of the fermion Compton wavelength),
the soliton acts as an inhomogeneous electric field that polarizes
the Dirac sea vacuum, thereby building up local fermion number at
the location of the kink. A straightforward one-loop computation
\cite{GW,W,WM} yields a fermion number taking arbitrary fractional
values:
\begin{equation}
N =-\frac{1}{\pi} \arctan{\(\frac{\hbar g\gamma}{m
c^2}\)}
\label{anyfrac}
\end{equation}

In our atomic system studied in Section IIB, we studied the
conjugation symmetric case where $\epsilon_k=0$ in \eq{EVP}, which
corresponds to the conjugation symmetric $m=0$ case in the continuum
limit. If instead we set a nonvanishing effective detuning $\delta$
for the hopping field between the two atomic levels, so that
$\epsilon_k=(-1)^k\delta/2$ with $\delta\neq0$, the symmetry between
the valence and the conduction band states is broken, and they are
no longer coupled together in a simple way. In the continuum limit
this corresponds to the $m\neq 0$ case in which the conjugation
symmetry is broken. Then the fractional part of the fermion number
may exhibit any value, in the continuum limit according to
\begin{equation}
N=-\frac{1}{\pi} \arctan{\(\frac{4\mu}{\delta}\)}\,.
\label{0kink}
\end{equation}
The ratio $4\mu/\delta$ between the coupling strength and the
effective detuning therefore determines the fractional part of the
particle number. In experiments this  could be engineered
accurately, allowing a potentially controlled way of preparing the
fractional part of the eigenvalues by tuning the mass of the Dirac
fermions.

An interesting further variation
would be to study the temperature dependence of the fermion number.
The above discussion is all for zero temperature, but at nonzero
temperature the fermion number may develop nontopological
contributions \cite{AD} and the fluctuations may become nonvanishing
\cite{DR}, depending on the form of the kink. Nonetheless, the fractional
fermion is not a singular feature of the theory at $T=0$ that vanishes
without a trace at any nonzero temperature. As a concrete example,
for a kink the finite temperature induced fermion number is
\cite{niemi,soni,parwani,AD,DR}
\begin{eqnarray}
\langle N\rangle_T=-\frac{2}{\pi}\sum_{n=0}^\infty \frac{y^2 \,
\sin\,\theta}{\left((2n+1)^2+y^2\right)\sqrt{(2n+1)^2\cos^2\,\theta
+y^2}}
\nonumber
\end{eqnarray}
where $\theta={\rm arctan}(4\mu/\delta)$, and $y=mc^2/(\pi T)$. This
expression smoothly reduces to the
zero temperature result (\ref{0kink}) as $T\to 0$. The fluctuations
$(\Delta N)^2\equiv \langle N^2\rangle-\langle N\rangle^2$ also has a
characteristic temperature dependence \cite{DR}.

The crucial part of our proposal for fractional particle number is
the em field $\varphi(x)$ in Eq.~{(\ref{defecthop})}. This is very
different from the fermion particle number fractionalization in
polymers, as our fermionic fields are not coupled to a bosonic
matter field with a domain-wall soliton. Instead, the coherent em
field with a topologically appropriate phase profile is coupled to
the FD atoms via internal transitions. This results in the
quantization of the FD atomic gas with nontrivial topological
quantum numbers corresponding to the soliton sector of the
relativistic 1+1 quantum field theory models of fractionalization. On
the other hand, a spatially constant phase profile $\varphi(x)$
represents the FD vacuum sector exhibiting integer particle numbers.


\section{Optical detection}\label{optsec}

The dimerized optical lattice resulting from the alternating pattern
of the hopping matrix elements causes the single-particle density of
states to acquire an energy gap, which  in the limit
$N_s\rightarrow\infty$ equals $4\hbar |\mu|$; see \eq{quasip}. The
zero state is located at the center of the gap. The resulting
excitations at half the gap energy could be detected by resonance
spectroscopy \cite{zero}. This provides indirect evidence of
fractionalization, as in the polymer systems~\cite{HEE88}. Because
in our scheme~\cite{RUO02} the gap is proportional to the amplitude
of the em field inducing the hopping, the size of the energy gap can
be controlled experimentally.

Such mid-gap spectroscopy, however, would not provide information
about fermion numbers. In this section we consider direct optical
measurements of the expectation value and of the fluctuations of the
fermion number in order to ascertain if they are compatible with
fractionalization.

We assume that far off-resonant light excites the atoms,
and consider the 1D optical lattice to be optically thin. We
take the light scatterer at each lattice site to be much smaller
than the wavelength of the detection light. In the case of off-resonant
excitation the amplitude of the light scattered from an essentially point
source is proportional to the number of atoms~\cite{JAV95}. Also, for
off-resonant excitation the scattering is coherent; the
scattered light has a fixed phase relation to the incoming light, and
is fully capable of interference. Thus, in a given point of observation
the positive frequency parts of the field operator for the light scattered
from the lattice sites simply sum up to
\begin{equation}
\hat{E}^+ = C\sum_k \alpha_kc^\dagger_kc_k\,.\label{lightnumber}
\end{equation}
$C$ is a constant containing the overall intensity scale of
the detection light.
The factors $\alpha_k$ include aspects such as intensity and phase
profiles of the detection light, effects of the spin state at each
site $k$ on light-atom coupling, and changes of the amplitude and
phase of the light as it propagates from the lattice sites to the
point of observation. A more detailed description how
\eq{lightnumber} is obtained can be found in Ref.~\cite{JAV95}.

\subsection{Measuring atom numbers}
\label{ATNUMMEAS}
In forward scattering and variations thereof such as phase contrast
imaging, the scattered and the incoming light interfere. The
ultimate measurement of the intensity in effect records the
expectation value of the scattered electric field $E=\langle
\hat{E}^+\rangle$. The observable at the detector is
\begin{equation}
E = \sum_k \alpha_k \langle c^\dagger_k c_k\rangle = \sum_{k p}
\alpha_k U^2_{kp} n_p\,.
\end{equation}
This is a linear combination of the expectation values of the
numbers of fermions at each lattice site with the coefficients
$\alpha_k$, exactly as introduced in~\eq{fracfermion}.

We now construct a numerical example about forward
scattering as a means to detect the fractional fermion. We make use of
the fact that the fermion species at the alternating lattice sites are
likely to be different. We assume that the detection light is far
blue-detuned in one species and far red-detuned in the other, and
that the two dipole matrix elements are comparable. One may then
find a laser tuning such that the intensity of the scattered light is
the same for both species. However, the lights scattered by the two
species are out of phase by $\pi$, and out of phase with the
incident light by $\pm\pi/2$. With the usual tricks of phase contrast
imaging, the relative phase of incident and scattered light is then
adjusted so that in interference the light from one species
directly adds to the incident light, and the light from the other
species subtracts. We incorporate the alternating sign into the
definition of our observable, and write
\begin{equation}
{E} = \sum_k \alpha_k (-1)^k \langle c^\dagger_k c_k\rangle \,.
\end{equation}

We take it that the incoming light is a plane wave with a constant
phase and amplitude. It will therefore just contribute a common
constant into the envelope factors $\alpha_k$, hence $\alpha_k$
basically stands for the amplitude of the light as transmitted
through the imaging system from the source point, lattice site ${\bf
r}_k$, to the observation point, ${\bf r}$. We adopt a rudimentary
physical-optics model according to which an imaging lense first
takes a Fourier transform of the light field at the object plane,
passes only the Fourier components that make it through the
aperture of the lense, and then takes another Fourier transform to
form the image.

Let us say that the geometry has been
arranged in such a way that all Fourier components of light in the plane
of the aperture up to the absolute value $K$ are passed, the rest are
blocked. Depending on where the object and image planes are, there might
be a scaling of the image with respect to the source, but we ignore both
this and the usual inversion of the image with respect to the object. The
transfer function of the lense from point
${\bf r}_k$ to the point
${\bf r}$, normalized to the peak value of unity, is then $2J_1(K|{\bf
r}-{\bf r}_k|)/(K|{\bf r}-{\bf r}_k|)$, and so we have the transmitted
field
\begin{equation}
{E}({\bf r}) = \sum_k (-1)^k \langle c^\dagger_k c_k\rangle\,\frac{2 J_1(K
|{\bf r}-{\bf r}_k|)}{K|{\bf r}-{\bf r}_k|}\,.\label{optics}
\end{equation}
\begin{figure}
\includegraphics[width=7cm]{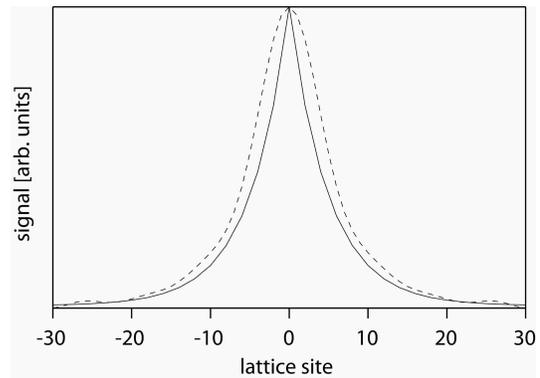}
\vspace{-10pt} \caption{Optical image (dashed line) of the
half-fermion density profile on top of a uniform background density
with the average number of 0.5 atoms per lattice site produced by a
specific imaging system, as discussed in the text. The density
profile (solid line) that is being imaged is also shown. The size of
the soliton is set by the choices $N_s=129$, $N_f=65$, and
$\mu=-0.1\,a$, corresponding to the correlation length of
$\xi\simeq10$ lattice sites.} \label{IMAGE}
\end{figure}

We present an example in Fig.~\ref{IMAGE}. We choose the parameters
$\mu = -0.1\,a$, and the numbers of sites and fermions $N_s=129$ and
$N_f=65$. Given the imaging light with the wavelength $\lambda$, we
use the cutoff wave number $K=2\pi/(\sqrt{5}\lambda)$. This would be
appropriate, for instance, if the imaging system had the numerical
aperture $F=1$ and the lattice resided near the focal plane of the
imaging lense; or if the numerical aperture were the rather extreme
$F=0.5$, and both the optical lattice and its image were removed by
two focal lengths from the lense. We assume that the wavelength of
the lattice light and of the imaging light are the same and that the
spacing between the lattice sites is $d=\lambda/4$. Under these
circumstances it would difficult to resolve individual lattice
sites. In practice a 1D optical lattice can be significantly
stretched by changing the intersection angle between the laser
beams, and so the resolution of the optical detection is not
necessarily limited by the lattice site spacing even if the
wavelengths of the lattice light and the light used for optical
detection were comparable. We set $d=\lambda/4$ simply to limit the
number of parameters to consider. We plot the optically imaged
fermion lattice along the line of the atoms (dashed line), and the
number of fermionic atoms in excess of the average occupation number
\half\ for the even-numbered sites that carry the lump with
$\<\tilde N\> =\half$ (solid line) as obtained from
Eq.~(\ref{fracfermion}). The curves are normalized so that the
maxima overlap.

Although the quantum operator ${\hat E}^+$ describing the phase contrast
imaging is not directly proportional to the fermion number operator
$\tilde N$ defined by \eq{fracfermion}, its expectation value produced by
the imaging system, \eq{optics}, nevertheless accurately depicts a
resolution rounded version of the half-fermion lump. In fact, phase
contrast imaging has been used for nondestructive
monitoring of a Bose-Einstein condensate~\cite{AND96}, and the
absorption of a single trapped ion has been detected
experimentally long ago~\cite{ION80}. While a lot of assumptions
went into our specific example, a light scattering experiment along these
lines should be feasible with the technology available today.

\subsection{Measuring atom number fluctuations}

\subsubsection{Intensity of Bragg peaks}

In the absence of interference with the incoming light, the
straightforward observable is the intensity of the light scattered from
the atoms. We have
\begin{equation}
I=\langle \hat{E}^-\hat{E}^+ \rangle\,.
\label{IVSE}
\end{equation}

With illumination of the optical lattice by a focused laser beam it is
possible to arrange the field strength of
the detection light vary from site to site approximately as a Gaussian.
Let us further assume, contrary to the previous section, that the atoms
at the lattice sites are identical as far as light scattering is
considered. If now the detection is carried out in the direction of
constructive interference (Bragg scattering) so that the light amplitudes
scattered from the lattice sites have the same phase, and far
enough from the lattice so that the propagation distance of light from
each lattice site to the detector is approximately the same, the field
strength is of the form
\begin{equation}
\hat{E}^+ =\sum_k e^{-(k/w)^2}  c^\dagger_k c_k\,.
\label{GWS}
\end{equation}
A measurement of the intensity of the light therefore measures the
square of the atom number operator, $I=\<\h N^\dagger \h N \>= |\<
\h N \>|^2 + (\Delta N)^2$, as in~\eq{FRS}, with the Gaussian
weights $\alpha_k^G$.

The key point in the detection of the correlated atom number
fluctuations responsible for the half-fermion is to measure the
coherent fluctuations \eqref{FRS}, i.e., to rely on interference of
light scattered from different lattice sites. If the light scattered
from individual sites can be resolved or if a too broad angular
average in the detection or other such cause wipes out the
interferences
[$\alpha^*_k\alpha_l\rightarrow\delta_{kl}\alpha^*_k\alpha_k$], the
scattered light probes the ``incoherent" fluctuations $(\Delta N)_i$
[\eq{incfluc}], as if the fermion number fluctuations in each site
were independent.

\begin{figure}
\includegraphics[width=7.0cm]{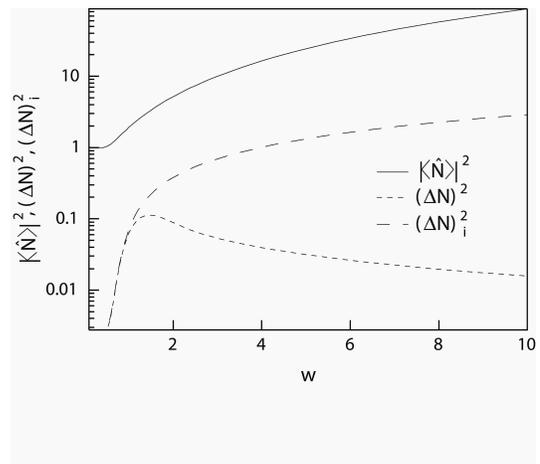}
\vspace{-10pt} \caption{The intensity of light scattered from the
optical lattice if the fermion numbers did not fluctuate, $|\<\h N
\>|^2$, and the additional intensity due to fermion number
fluctuations $(\Delta N)^2$, as a function of the size of the focus
$w$ of the driving light given in the lattice units. We also display
the added intensity $(\Delta N)^2_i$ that would result if fermion
number fluctuations were uncorrelated between adjacent lattice
sites. The soliton parameters are $N_s=129$, $N_f=65$,
$\mu=-0.9\,a$.} \label{INTENSITY}
\end{figure}

The problem in detecting the fractional fermion number is that in
the measured signal of the light intensity the fluctuations $\Delta
N$ appear on a nonzero background. We demonstrate by plotting in
Fig.~\ref{INTENSITY} separately the contribution $|\<\h N \>|^2$, as
if the fermion numbers were precisely fixed, and the fluctuation
term $(\Delta N)^2$. We also show the fluctuations $(\Delta N)_i^2$
that would result if the fermion number fluctuations at adjacent
sites were uncorrelated. These are given as functions of the width
of the focus $w$ of the laser beam. Here $N_s=129$, $N_f=65$, and we
choose $\mu=-0.9\,a$ to make a sharply localized zero state.

In our discussion we employ the Gaussian width $w=4$,
which would ordinarily  represent very tight focusing. Here the
contribution from atom number fluctuations to light intensity is two
orders of magnitude below the coherent intensity, whereas the
fluctuations from uncorrelated fermion numbers would make a contribution
an order magnitude smaller than the coherent intensity. As our detection
light was assumed to be far off resonance, photon number fluctuations
are Poissonian. Under otherwise ideal conditions, including absolute
knowledge and control of all experimental parameters pertaining to
intensity of the scattered light, the detection of about a hundred
photons could reveal the difference between correlated and uncorrelated
fermion numbers, whereas a quantitative study of the actual correlated
fermion number requires the detection of about 10,000 photons.
Unfortunately, a large number of scattered photons means a large number
of recoil kicks on the fermions. Currently available optical lattices
likely cannot absorb the assault of hundreds of photon recoils without
developing some form of a dynamics that complicates the phenomena we are
analyzing.

\subsubsection{Variations of intensity measurements}

While the scattered light in principle conveys information
about the fluctuations of atom number, the method to extract the
information we have discussed so far is quite challenging.

A way to discard the uniform background of \half\, fermions per
lattice site, as in going from operator $\h N$ to $\tilde N$
in~\eq{fracfermion}, would be most useful. One might, for instance, think
of the alternating-sign Gaussian envelope $\alpha_k^A$, as
defined in \eq{agaussprof} and (in principle) realized as discussed in
Sec.~\ref{ATNUMMEAS}, and still look in the direction of ordinary Bragg
scattering. By virtue of the first form of \eq{CSUM}, the coherent
signal $|\<\h N
\>|^2$ from the background of \half\ indeed cancels in the limit of a
broad envelope. However, the fluctuation part
$(\Delta N)^2$ as appropriate for
$\alpha_k^A$ is not small, and does not directly bear witness to the
small fluctuations of the atom number of the half-fermion. On
the contrary, as discussed in Sec.~\ref{FLUSTU}, in this case the
fluctuations of the atom number under the Gaussian envelope are
larger that one would expect if the atom number fluctuations at
adjacent sites were uncorrelated.

A scheme that fares better is based on the observation that off-resonant
light scattering is coherent. If the atom numbers were fixed, the
field scattered by the lattice as a whole were also completely coherent
with the incoming light. One could then tap part of the incident
light and impose a suitable attenuation and phase on it so that on the
detector the incident light exactly cancels the scattered light.  Given
the fluctuating atom numbers, it is correspondingly possible in principle
to arrange things so that the detected intensity originates entirely from
atom number fluctuations.

In technical terms, assume that the incident and scattered fields are
superimposed on the detector so that the electric field operator is
\begin{equation}
\hat{E}^+ = C\sum_k \alpha_k(c^\dagger_k c_k - \langle c^\dagger_k
c_k\rangle)
\label{CANCELATION}
\end{equation}
instead of~\eq{lightnumber}. This eliminates the coherent part of
the intensity
  $\propto|\<\h N\>|^2$, leaving
only the fluctuation part $(\Delta N)^2$ in~\eq{FRS}. The situation
of~\eq{CANCELATION} may be realized operationally by  minimizing the
detected intensity with adjustments of the attenuation and the phase
of the canceling light. The remaining intensity is then directly
proportional to the the square of atom number fluctuations under the
envelope of the detection light. A reduction in the minimum detected
intensity with an increasing size of the focus of the driving light
would be a signature of anomalous atom number fluctuations
characteristic of the fractional fermion. Another approach to
eliminate the coherent part of the light intensity is to exploit the
diffraction pattern of the scattered light from the regular array of
the lattice sites and detect the photon number fluctuations at the
location of the destructive interference of the coherent
$\propto|\<\h N\>|^2$ contribution of the scattered light.

Although our goal here is not a specific experimental design,
another variation with potential to overcome the accumulation of
atom recoil merits a mention.  So far we have dealt with what in
essence is spontaneous Bragg scattering. Recently, induced Bragg
scattering has been introduced as a method to study the condensates
in detail~\cite{STA99,SAB05}. In optical lattices the strong spatial
confinement may complicate the Bragg spectroscopy measurements
\cite{mott1d,DU07}, but the advantage of the light-stimulated Bragg
scattering is that conceivably the light pulses could be made so
short that the harmful effects of photon recoil do not have time to
build up during the measurement.

\section{Concluding remarks}\label{concsec}

The optical lattice could be part of a larger harmonic trap, which
may also affect the energy profile $\epsilon_k$ in the Hamiltonian
\eqref{ham}. The length scale over which the energy variation due to
the trap becomes comparable to the energy gap is then $r_0\equiv
(8\hbar|\mu|/M\omega^2)^{1/2}$, where $\omega$ denotes the trap
frequency and $M$ the mass of the atom. For the system to have a
locally homogeneous gap value, this length scale should be much
larger than the correlation length $\xi$. Moreover, in order to have
a better experimental control over the atom numbers and to maintain
the one-half filling throughout the lattice, we require that the
energy gap is nonvanishing everywhere, or $r_0\agt N_h d$. The
nonvanishing gap parameter and the locally homogeneous limit could
therefore be reached with a sufficiently weak trap and by
experimentally controlling the size of the gap. By using
sufficiently large values of $|\mu|$, it might also be possible to
study the effective zero temperature limit $|\mu|\gg k_B T/\hbar$
and to have a better control over the light scattering measurements.
Finally, the sharp edges of the optical lattice system with
hard-wall boundaries could be prepared, e.g., by shining
blue-detuned laser beams at the lattice ends.

We could possibly also construct complex higher dimensional models
in 2D or 3D optical lattices with atoms coupled to nontrivial em
fields. In relativistic (2+1)D quantum field theory, a fermionic
field coupled to a bosonic field exhibiting a vortex profile results
in fermion particle number fractionalization
\cite{NIE86,SEM84,HAL88,PAR91,FLE91}. It may also be possible to
find analogies to relativistic (3+1)D quantum field theoretical
models \cite{JAC76}, e.g., by using em field superpositions to
prepare topologically nontrivial atomic configurations
\cite{RUO01,STO01,RUO03,SAV03,RUO05b,RUS05}. The significance of
localized zero energy modes in fermionic systems has also recently
been discussed in the context of quantum teleportation \cite{SEM06},
using results from the quantum field theoretical analysis in
\cite{NS84}.

\acknowledgments{This work was financially supported by the EPSRC, the DFG,
the U.S. DOE., the U.S. NSF, and NASA. GD thanks the University of
Adelaide and the ITP in Heidelberg for hospitality and support while
on sabbatical.}


\begin{thebibliography}{01234}
\bibitem{signcomment} We have here redefined some signs of the variables
used in Ref.~\cite{JAV03} in order to simplify the overall notation.
\bibitem{zero} In a finite-size lattice the midgap mode may also exist
in the vacuum state. However, this is always spatially separated
from the soliton bound state.
\end{thebibliography}
\end{document}